\documentclass[%
 reprint,
showpacs,
 amsmath,amssymb,
 aps,
prb,
floatfix
]{revtex4-1}
\usepackage{braket}
\usepackage{graphicx}
\usepackage{dcolumn}
\usepackage{bm}
\usepackage{color}

\begin{document}

\preprint{APS/123-QED}

\title{Topological band structure of surface acoustic waves on a periodically corrugated surface}

\author{Tomohiro Inoue$^1$ }

\author{Shuichi Murakami$^{1,2}$}%

\affiliation{%
$^1$Department of Physics, Tokyo Institute of Technology, 2-12-1 Ookayama, Meguro-ku, Tokyo, 152-8551,  Japan \\
$^2$TIES, Tokyo Institute of Technology, 2-12-1 Ookayama, Meguro-ku, Tokyo, 152-8551, Japan
}%

\date{\today}

\begin{abstract}
Surface acoustic waves (SAWs) are elastic waves localized on a surface of an elastic body. We theoretically study topological edge modes of SAWs for a corrugated surface. We introduce a corrugation forming a triangular lattice on the surface of an elastic body. We treat the corrugation as a perturbation, and construct eigenmodes on a corrugated surface by superposing those for the flat surface at wavevectors which are mutually different by reciprocal lattice vectors. We thereby show emergence of Dirac cones at the $K$ and $K'$ points analytically. Moreover, by breaking the time-reversal symmetry, we show that the Dirac cones open a gap, and that the Chern number for the lowest band has a nonzero value. It means existence of topological chiral edge modes of SAWs in the gap.
\end{abstract}

\pacs{Valid PACS appear here}
\maketitle


\section{\label{sec:level1}INTRODUCTION}
Various topological phases in electronic systems have been studied since the quantum Hall effect (QHE) was discovered in 1980~\cite{klitzing}. In the QHE, topologically protected chiral edge modes are formed while the bulk is an insulator in two-dimensional systems. 
This is attributed to a nontrivial topology of the band structure characterized by the Berry curvature. Effects of the Berry curvature appear in various Hall effects. 
Hall effects for other particles and quasi-particles have been studied following the studies in electron systems. Among them is the phonon Hall effect \cite{0953-8984-23-30-305402,PhysRevLett.105.225901,PhysRevB.86.104305} where a transverse heat current is induced by temperature gradient. Moreover, in recent years, topologically protected phonon systems are studied theoretically and experimentally~\cite{PhysRevApplied.10.014017,0022-3727-51-17-175302,PhysRevB.96.094106,0953-8984-30-22-225401,PhysRevLett.114.114301,0953-8984-30-34-345401,PhysRevB.98.014302,1367-2630-20-7-073032,8091821}, which are analogous to the QHE in the electronic systems.

In this paper we theoretically study topological chiral edge modes in the surface acoustic waves (SAWs). The SAWs are a special type of elastic waves localized on a surface of an elastic body and they decay exponentially into its bulk~\cite{landau}. The SAWs are drawing attention both for scientific interests and for applicational purposes~\cite{PhysRevB.80.014301,doi:10.1063/1.4870045,doi:10.1063/1.348969,PhysRevLett.119.077202,PhysRevLett.68.2464,doi:10.1063/1.4729891,PhysRevB.52.11475,doi:10.1063/1.3626853}. We focus on the SAWs, because its band structure can be designed by controlling surface corrugations. Dispersion relations of the SAWs for a periodically corrugated surface have already been studied~\cite{glass,GIOVANNINI1993783,1367-2630-13-1-013037,doi:10.1063/1.1290388,doi:10.1063/1.332038,PhysRevB.28.728}. For a surface with one-dimensional corrugation, it was shown that a gap opens at the boundary of the Brillouin zone~\cite{glass}. However, no analytical studies have been conducted to combine topological phenomena like QHE and the SAWs for the corrugated surface in the continuum theory of elasticity. 
There have been a number of studies on  topological phenomena  in spring-mass systems and various mechanical systems \cite{KaneLubensky,Paulose,Rocklin,
Rocklin-PRL,Stenull,Nash-PNAS,
Bilal,Wang-PRL,Mousavi,Susstrunk,Pal,Kariyado,Wang-NJP,Brendel-PNAS,Chaunsali-PRL,Chaunsali-PRB,Prodan,yu2016surface,wang2018guiding,ma2018edge,doubleWeyl,negative}.
Among them, there are only a few previous works on topological phenomena on SAWs. For example, quantum valley Hall effect is proposed in Ref.~\onlinecite{wang2018guiding} and a ``phononic graphene'' has been realized experimentally  in Ref.~\onlinecite{yu2016surface}. From an analogy with topological semimetals, a Weyl phononic crystal was fabricated \cite{negative}. Its novel topological surface modes, analogous to Fermi-arc surface states in a Weyl semimetal, were observed and its negative refraction was measured \cite{negative}. It can be regarded as topological surface modes of SAWs in Weyl phononic crystals.

In this paper, we study SAWs flowing along a surface with a two-dimensional corrugation forming a triangular lattice and their topological bands. From the viewpoint of symmetry, Dirac cones are expected to appear at the $K$ and $K'$ points, which are the vertices of the Brillouin zone, as seen in similar hexagonal systems~\cite{yu2016surface,PhysRevB.89.134302,PhysRevLett.108.174301,doi:10.1063/1.4998438,doi:10.1063/1.5004073}. First of all, we analytically show emergence of the Dirac cones for the SAWs. Furthermore, the bands of the SAWs are expected to be topological  when the Dirac cones open a gap, in analogy with those in electronic systems. Indeed, we show emergence of topological chiral edge modes within the gap, analogous to those in the QHE, by introducing a term which breaks time-reversal symmetry. This phase has chiral edge modes on the surface of an elastic body. In other words, we can create a one-dimensional elastic wave that is topologically protected along the edge of the two-dimensional surface.

Throughout this paper, the corrugation is treated as a small perturbation. This enables us to study the eigenmodes, dispersions, and Berry curvatures in an analytic way. The analytic results help us to obtain physical insights into the physics of topological modes of the SAWs. We note that even when the corrugation becomes larger, the topological nature persists as long as the band gap remains open. 

This paper is organized as follows. Section~\ref{sec:level2} is devoted to showing the emergence of the Dirac cones at the $K$ and $K'$ points on the corrugated surface. In Sec.~\ref{sec:level3}, we show topological nature of the band structure for SAWs by breaking the time-reversal symmetry, and we conclude this paper in Sec.~\ref{sec:level4}.
\section{\label{sec:level2}EMERGENT DIRAC CONES OF SAWS AROUND $K$ AND $K'$ POINTS ON THE CORRUGATED SURFACE}
In this section, we show that Dirac cones appear at $K$ and $K'$ points in the Brillouin zone for SAWs on a corrugated surface forming a triangular lattice. For this purpose, firstly, we calculate eigenfrequencies and eigenmodes at the $K$ and $K'$ points. After that, we show that the Dirac cones appear around the $K$ and $K'$ points by the $\mbox{\boldmath $k$}\cdot\mbox{\boldmath $p$}$ perturbation theory using the eigenmodes at the $K$ and $K'$ points.
\begin{figure}[t]
\includegraphics[width=8.0cm]{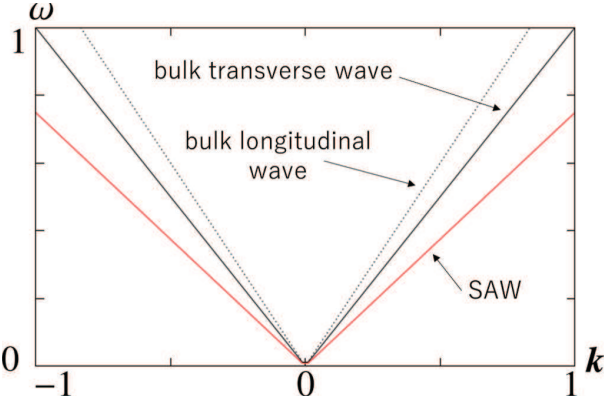}
\caption{\label{fig:dispersion_flat} Dispersion relations of SAWs for the flat surface. The black line represents the bulk transverse wave mode $\omega=c_t k$ and the dashed black line represents the bulk longitudinal wave mode $\omega=c_l k$. The red line represents the dispersion relation of the SAWs $\omega=c_R k$. Here, we set $c_t = 1$ and $c_l = 1.2$.}
\end{figure}
\subsection{\label{sec:level2 subsec:level1}Preliminaries: Acoustic Waves in the Bulk and on the Flat Surface}
In this section, we review acoustic waves in the bulk and on the flat surface of a microscopically isotropic elastic body following Ref.~\onlinecite{landau}, while systems with microscopic anisotropy would be of interest as a future work. We begin with the classical equation of motion for the elastic body:  $\rho \ddot{u}_i =\sum_j \partial \sigma_{ij}/\partial x_j$. Here $\rho$ is the density of the elastic body, $u_i$ is the $i$th component of the displacement vector $\mbox{\boldmath $u$}$, $\sigma_{ij}$ represents the stress tensor and $x_j$ represents the coordinates $(x_1,x_2,x_3)=(x,y,z)$. Force balance implies that this stress tensor $\sigma_{ij}$ is given in terms of the strain tensor $u_{ij}$ as $\sigma_{ij}=\rho(c_l^2 - c_t^2)\sum_{l} u_{ll} \delta_{ij} + 2\rho c_t^2 ( u_{ij} -\sum_{l} u_{ll} \delta_{ij}/3 )$~\cite{landau}, where $c_l$ and $c_t$ represent the velocities of the longitudinal and transverse waves, respectively, and $u_{ij}=(\mathrm{d}u_i/\mathrm{d}x_j+\mathrm{d}u_j/\mathrm{d}x_i)/2$ is the strain tensor. Therefore, the equation of motion can be written only in terms of the displacement vector:
\begin{eqnarray}
\ddot{\mbox{\boldmath $u$}}=c_t^2 \mbox{\boldmath $\nabla$}^2\mbox{\boldmath $u$}+(c_l^2 - c_t^2)\mbox{\boldmath $\nabla$}(\mbox{\boldmath $\nabla$}\cdot\mbox{\boldmath $u$}).
\label{EOM}
\end{eqnarray}
Then we obtain wave equations of the longitudinal wave: $\partial^2 \bm{u}_l/\partial t^2=c_l^2 \bm{\nabla}^2\bm{u}_l$ and the transverse wave: $\partial^2 \bm{u}_t/\partial t^2=c_t^2\bm{\nabla}^2\bm{u}_t$ with $\bm{\nabla} \times \bm{u}_l=0$ and $\bm{\nabla} \cdot \bm{u}_t=0$ from Eq.~(\ref{EOM}). To calculate dispersions of the SAWs for a surface of the elastic body occupying the region $z<0$, we combine solutions of the transverse and longitudinal waves localized near the surface $z=0$ as follows:
\begin{align}
&u_x=\frac{k_x}{k} \left(A_le^{\alpha_l z} + A_t e^{\alpha_t z} \right) e^{i(\bm{k}\cdot \bm{r} - \omega t)} ,\label{u_x flat}\\
&u_y=\frac{k_y}{k }\left(A_le^{\alpha_l z} + A_t e^{\alpha_t z} \right) e^{i(\bm{k}\cdot \bm{r} - \omega t)} ,\label{u_y flat}\\
&u_z=\left(-\frac{\alpha_l}{k} A_le^{\alpha_l z} - \frac{k}{\alpha_t} A_t e^{\alpha_t z} \right)i e^{i(\bm{k}\cdot \bm{r} - \omega t)} ,\label{u_z flat}
\end{align} 
where $\alpha_l\equiv \sqrt{k^2 - \omega^2/c_l^2}$, $\alpha_t\equiv \sqrt{k^2 - \omega^2/c_t^2}$, $\bm{r}=(x,y)$, $\bm{k}=(k_x,k_y)$ is the wavevector along the surface $z=0$, and $\omega$ is the frequency. $A_l$ and $A_t$ are amplitudes of the longitudinal and transverse waves respectively. To obtain dispersions of SAWs, we impose boundary conditions that the surface is stress-free. In terms of the stress tensor $\sigma_{ij}$ and the unit vector $\mbox{\boldmath $e$} = (e_1,e_2,e_3)$ that is normal to the surface, the boundary conditions are written as $\sum_{j} \sigma_{ij} e_j |_{z = 0}=0\ (i = 1,2,3) $. By writing $\sigma_{ij}$ in terms of the displacement vector $(u_x,u_y,u_z)$ given by Eqs.~(\ref{u_x flat})-(\ref{u_z flat}) and by putting $\bm{e}=(0,0,1)$, we obtain dispersions of the SAWs $\omega =c_t \xi k$ and the ratio of the coefficients $A_t=(-1+\xi^2/2)A_l$, where $\xi$ is a solution of the
equation $\xi^6 -8\xi^4 + 8\xi^2(3-2c_t^2/c_l^2) -16(1-c_t^2/c_l^2)=0$ in the range of $\xi <1$~\cite{landau}. This surface wave is called the Rayleigh wave, and its velocity is given by $c_R \equiv c_t \xi$. We show the dispersion of the SAW in Fig.~\ref{fig:dispersion_flat} for $c_t=1$ and $c_l=1.2$ as an example.
\subsection{\label{sec:level2 subsec:level2}Dispersion Relations of SAWs at $K$ point}
\begin{figure*}
\includegraphics[width=14.0cm]{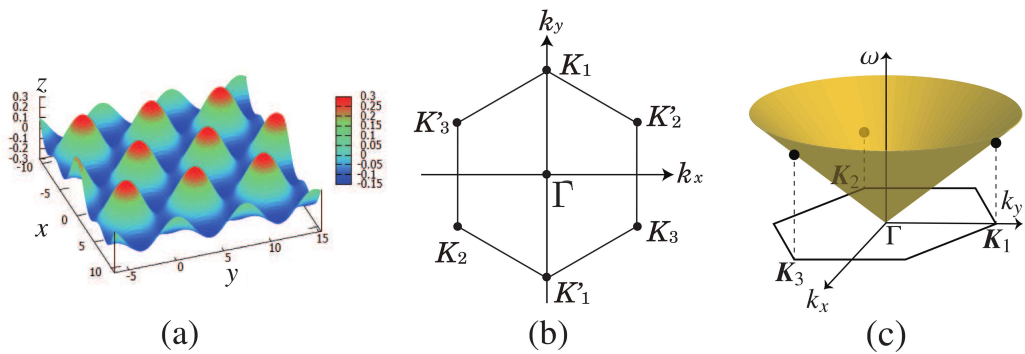}
\caption{\label{fig:surface_and_Kpoint} Corrugated surface with a triangular lattice, and calculation of its band structure. (a) Profile of the surface corrugation of the elastic body described by Eq.~(\ref{zeta}). The figure shows $z=\zeta(x,y)$ with the values of the parameters $d = 0.1$ and $a=4\pi/\sqrt{3}$. (b) Brillouin zone of the model. We call the six corners $\bm{K}_j$ and $\bm{K}'_j$
($j=1,2,3$). As reciprocal lattice vectors, the three points $\bm{K}_j$ are equivalent, meaning that
their differences are reciprocal lattice vectors, and are called $K$ point in the standard notation for special wavevectors in the Brillouin zone. 
Likewise, the three points $\bm{K}'_j$ are equivalent and are called $K'$ point, but $K$ and $K'$ are not equivalent. The $\Gamma$
point denotes the origin $\bm{k}=(0,0)$. 
(c) Schematic picture of the dispersion relation of SAWs when the surface is flat. It is shown in the first Brillouin zone for the triangular lattice. The dispersion relation of SAWs for the flat surface has a conical structure whose apex is at the $\Gamma$ point denoting the center of the first Brillouin zone. In the calculation of the dispersion of SAWs around $\bm{K}_1$ for the corrugated surface, we consider only  hybridization of the waves near the three $K$ points $\bm{K}_1,\bm{K}_2$ and $\bm{K}_3$.}
\end{figure*}

In this paper, we consider a microscopically isotropic elastic body whose surface is corrugated periodically, forming a triangular lattice. We set the profile of the surface corrugation along the $xy$ plane as
\begin{eqnarray}
z&=&\zeta(x,y) \nonumber \\
&\equiv&d( \cos((\bm{b}_1+\bm{b}_2)\cdot \bm{r}) + \cos(\mbox{\boldmath $b$}_1\cdot \mbox{\boldmath $r$}) + \cos(\mbox{\boldmath $b$}_2\cdot \mbox{\boldmath $r$}))
\label{zeta}
\end{eqnarray}
instead of $z=0$ (Fig.~\ref{fig:surface_and_Kpoint} (a)).
 The elastic body exists in the region $z\leq \zeta(x,y)$, and the region $z>\zeta(x,y)$ is a vacuum. Here, we set $\mbox{\boldmath $b$}_1=(2\pi/a)(1/\sqrt{3},1),\ \mbox{\boldmath $b$}_2=(2\pi/a)(1/\sqrt{3},-1)$, and $d,a $ are positive constants. This surface corrugation forms a triangular lattice with a lattice constant $a$ and primitive reciprocal lattice vectors $\mbox{\boldmath $b$}_1$ and $ \mbox{\boldmath $b$}_2$. 
This system has sixfold rotation and time-reversal symmetries. 
The reason of choosing the triangular lattice, instead of the honeycomb lattice adopted often 
in previous works on topological bands \cite{Haldane,Wang-PRL,Wang-NJP,0022-3727-51-17-175302} is because in the present continuum 
model, the number of Fourier components to realize surface corrugation is smaller
for the triangular lattice (see Eq.~(\ref{zeta})), making the theory simpler.
In the following  we assume $d\ll a$ so that the corrugation $\zeta$ can be treated perturbatively. 

One can  construct solutions for eigenmodes of SAWs on the corrugated surface perturbatively from those on the flat surface. 
Because the calculation is lengthy, we describe its outline here, with its details given in Appendix~\ref{eigenvalues and eigenvectors of SAW}.
First, we point out that the corrugation hybridizes the plane-wave solutions on the flat surface with various wavevectors which are different from each other by the reciprocal lattice vectors of the surface. This method is similar to that applied to a one-dimensional plasmonic crystal for surface plasmons~\cite{kitamura2013hermitian}. Then we obtain solutions of Eq.~(\ref{EOM}) in the region $z \leq \zeta$ (see Eqs.~(\ref{u_x})-(\ref{u_z}) in Appendix~\ref{eigenvalues and eigenvectors of SAW}). 
To determine dispersions and the coefficients of plane-wave solutions, we impose the stress-free boundary conditions. Here, the equations of boundary conditions can be regarded as infinite-dimensional matrix equations because an infinite number of plane waves are involved in the equations. 
In order to solve them analytically, we focus on the $K$ point in the Brillouin zone. 
In the lowest order in the perturbation theory at the $K$ point, we only have to consider the terms only from $\bm{K}_1=4\pi/(3a)(0,1)$, $\bm{K}_2=4\pi/(3a)(-\sqrt{3}/2,-1/2)$ and $\bm{K}_3=4\pi/(3a)(\sqrt{3}/2,-1/2)$, and we can ignore contribution from other $\bm{k}$ points (Fig.~\ref{fig:surface_and_Kpoint}). In addition, we approximate the ratio of the longitudinal amplitude to the transverse amplitude to be the value at the zeroth order of the perturbation because $\zeta$ is small~\cite{glass}. In other words, we approximate the ratio to be equal to that for a flat surface. By using the above approximations for boundary conditions, we can obtain the eigenvectors
\begin{eqnarray}
&&\bm{A}_l^{(1)}=(1,\ 1,\ 1) \ , \label{A_l^1}\\
&&\bm{A}_l^{(2)}=(1,\ \eta,\ \eta^{2}) \ , \label{A_l^2}\\
&&\bm{A}_l^{(3)}=(1,\ \eta^{2},\ \eta) \ , \label{A_l^3}
\end{eqnarray}
where $\eta=e^{2\pi i/3}$ and the $i$-th components of $\bm{A}_l^{(n)}$, $A_{l,i}^{(n)}\ (i=1,2,3)$, correspond to the $\bm{K}_i$ points and the superscripts $(n)\ (n=1,2,3)$ are labels for the eigenmodes. The corresponding eigenvalues are obtained as
\begin{eqnarray}
\omega_K^{(1)}&=&\sqrt{\frac{\xi^4 + 2d\beta (\beta \beta_{\lambda} - \nu^2)K}{\xi^2 -2d\beta(\beta \beta_{\lambda} + \nu)K}}c_t K, \\
\omega_K^{(2)}=\omega_K^{(3)} &=& \sqrt{\frac{\xi^4 -d\beta (\beta \beta_{\lambda} - \nu^2)K}{\xi^2 +d\beta(\beta \beta_{\lambda} + \nu)K}}c_t K \label{omega_K^(D)}
\end{eqnarray}
at the $K$ point, where $K=4\pi/(3a),\ \beta=\sqrt{1-\xi^2},\ \beta_{\lambda}=\sqrt{1-\xi^2\lambda^2}$, $\nu=-1+\xi^2/2$ and $\lambda=c_t/c_l$ (the details are in Appendix~\ref{eigenvalues and eigenvectors of SAW}). Since the eigenmodes with $n = 2$ and $n=3$ are degenerate and form a doublet, we write $\omega_K^{(D)} \equiv \omega_K^{(2)}=\omega_K^{(3)}$. Thus, the triply degenerate modes at the $K$ point on the flat surface are lifted to a singlet ($\omega_K^{(1)}$) and a doublet ($\omega_K^{(2)}$ and $\omega_K^{(3)}$).

\subsection{\label{sec:level2 subsec:level3}Solutions away from the $K$ Point}
In Sec.~\ref{sec:level2 subsec:level2} we obtained the eigenmodes at the $ K $ point. In this section, using these eigenmodes at the $K$ point, we construct eigenmodes away from the $ K $ point by the $\mbox{\boldmath $k$}\cdot\mbox{\boldmath $p$}$ perturbation theory. Before applying the $\bm{k}\cdot \bm{p}$ perturbation theory we note that, according to Eq.~(\ref{EOM}), the equation of motion is non-Hermitian: 
\begin{eqnarray}
\omega \psi = \tilde{H} \psi,\  
\tilde{H}=\begin{pmatrix}
0 & iI_3 \\
-ih'_3 & 0
\end{pmatrix}
,
\label{eq:nonHerm}
\end{eqnarray}
where $\psi=(\mbox{\boldmath $u$}, \mbox{\boldmath $v$})^T$, with $\mbox{\boldmath $v$}=\dot{\mbox{\boldmath $u$}}$ representing the velocity.
In the matrix $\tilde{H}$, $h'_3$ is a $3\times3$ matrix with its $(i,j)$ component given by $(h'_3)_{ij}=-c_t^2 \delta_{ij}\mbox{\boldmath $\nabla$}^2 -(c_l^2 - c_t^2)\nabla_i \nabla_j$, and $I_3$ is the $3\times3$ identity matrix. This non-Hermitian form of the matrix $\tilde{H}$ is inconvenient for formulating the Berry curvature in this setup. Hence, we transform the non-Hermitian eigenvalue problem
(\ref{eq:nonHerm}) into a generalized Hermitian eigenvalue problem; this guarantees fundamental properties of the Berry curvature,
such as gauge invariance, which in turn leads to appearance of the Berry curvature in various physical phenomena.

In order to make the problem Hermitian, we introduce a new Hermitian matrix $\gamma$, so that the eigenvalue equation is rewritten as $\gamma \tilde{H} \psi=\omega \gamma \psi$ with $\gamma \tilde{H}$ being Hermitian. It is not trivial whether such a Hermitian matrix $\gamma$ exists, which makes $\gamma \tilde{H}$ to be also Hermitian. In the present case, to deduce the form of the Hermitian matrix $\gamma$, we focus on a conserved quantity in this equation. Using the equation expressed in terms of the time derivative $H\psi=i\gamma \partial \psi /\partial t$ with $H \equiv \gamma \tilde{H}$ being Hermitian, we find that $\psi^{\dag}H\psi$ is the conserved quantity, i.e., $\mathrm{d}(\int\mathrm{d}V\psi^{\dag}H\psi)/\mathrm{d}t=0$. Therefore, we identify $\psi^{\dag}H\psi$ with the energy density of the elastic body:
\begin{eqnarray}
\psi^{\dag}H\psi \propto \mathcal{E}, \ \mathcal{E}=\frac{1}{2}\rho \dot{\mbox{\boldmath $u$}}^2 + \frac{1}{2} \sum_{ij}\sigma_{ij} u_{ij}.
\end{eqnarray}
Since the stress tensor $\sigma_{ij}$ and the strain tensor $u_{ij}$ can be expressed in terms of the displacement vector $\mbox{\boldmath $u$}$, the energy density $\mathcal{E}$ can be rewritten in terms of $\psi$. By ignoring surface terms, we can rewrite the expression of $\mathcal{E}$:
\begin{eqnarray}
\mathcal{E}=\frac{1}{4}\rho \psi^{\dag} 
\begin{pmatrix}
h'_3 &0 \\
0&I_3
\end{pmatrix}
\psi\ .
\end{eqnarray}
Therefore, using the equation of motion $h'_3 \mbox{\boldmath $u$}=-\ddot{\mbox{\boldmath $u$}}$, we obtain 
\begin{eqnarray}
&&H\psi=\omega\gamma \psi, 
\label{eigen eq}\\
&&H=\gamma\tilde{H}=
\begin{pmatrix}
h'_3 &0 \\
0&I_3
\end{pmatrix}
,\ 
\gamma=
\begin{pmatrix}
0 &iI_3 \\
-iI_3&0
\end{pmatrix}
\end{eqnarray}
as a new Hermitian form of the equation of motion. It is a generalized Hermitian eigenvalue equation. The norm of the wavefunction is defined as $N=\int_{\rm unit\ cell} \mathrm{d}V \psi^{\dag}\gamma \psi$. Here we adopt the eigenmodes given by Eqs.~(\ref{A_l^1})-(\ref{A_l^3}) and their norms are calculated as
\begin{eqnarray}
N=\frac{9\sqrt{3}}{4\pi} a^3 \omega \left( \frac{\beta^2 + 1}{2\beta} + \frac{\beta_{\lambda}^2 +1}{2\beta_{\lambda}^3}\nu^2 + \frac{2\nu}{\beta + \beta_{\lambda}} \right),
\label{N}
\end{eqnarray}
where $\omega$ denotes $\omega_K^{(1)}$ or $\omega_K^{(D)}(=\omega_K^{(2)}=\omega_K^{(3)})$ depending on the eigenmodes used in calculating the norm.

We note 
that in Ref.~\onlinecite{Susstrunk} a similar transformation from a non-Hermitian eigenvalue
problem for phonons in a spring-mass model to a generalized Hermitian 
eigenvalue problem is developed.
The formalism in Ref.~\onlinecite{Susstrunk} is basically limited to models with 
a discrete degree of freedom within the unit cell, such as spring-mass models. In contrast, our formalism here gives a recipe for general 
phononic systems, including even continuum systems, and it includes the formalism in Ref.~\onlinecite{Susstrunk} as a special case. We note that 
this extension to continuum systems is nontrivial, because of the infinite number of variables within the unit cell. 
Namely, in applying the method in Ref.~\onlinecite{Susstrunk} to the present case, we need to calculate the square root of the operator $h'_3$, and it is technically difficult. Thus, for continuum systems, only our method is applicable for making the eigenvalue problem Hermitian.

Thus far, we have obtained the Hermitian eigenvalue equation (\ref{eigen eq}). Using the eigenmodes at the $K$ point obtained in Sec.~IIB, we construct eigenmodes at the points away from the $ K $ point by the $\mbox{\boldmath $k$}\cdot \mbox{\boldmath $p$}$ perturbation theory. For that purpose, we express the displacement vector in the Bloch form: $\mbox{\boldmath $u$}=\mbox{\boldmath $U$}e^{i(\bm{k} \cdot \bm{r}-\omega t)}$. Accordingly, the wavefunction  $\psi=(\bm{u},\bm{v})^T$ is rewritten as $\psi=\Psi e^{i(\bm{k} \cdot \bm{r}-\omega t)}$. From
\begin{eqnarray}
(h'_3\mbox{\boldmath $u$})_i&=&\sum_j[-(c_l^2 - c_t^2)(\nabla_i+ik_{i})(\nabla_j+ik_{j}) \nonumber \\
&\ \ & - c_t^2(\mbox{\boldmath $\nabla$} + i\mbox{\boldmath $k$})^2 \delta_{ij}]U_j e^{i(\bm{k} \cdot \bm{r}-\omega t)},
\end{eqnarray}
we rewrite Eq.~(\ref{eigen eq}) as
\begin{eqnarray}
H_0\ket{\Psi}=\omega \gamma \ket{\Psi}\ ,\ \ H_0=\begin{pmatrix}
h_3(\bm{k})&0 \\
0&I_3
\end{pmatrix}
\label{hermitian equation for kp}
\end{eqnarray}
where $(h_3(\bm{k}))_{ij}=-(c_l^2 - c_t^2)(\nabla_i + ik_{i})(\nabla_j + ik_{j})-c_t^2 (\mbox{\boldmath $\nabla$} + i\mbox{\boldmath $k$})^2 \delta_{ij}$. Here, we have introduced the bra-ket notation for $\Psi$, and the norm of $\Psi$ is given by $N=\braket{\Psi_n(\bm{k})| \gamma |\Psi_n(\bm{k})}\equiv \int_{\rm unit cell}dV\Psi_n(\bm{k})^{\dagger} \gamma \Psi_n(\bm{k})$. First we set $\mbox{\boldmath $k$}=\bm{K}_1$, and we write Eq.~(\ref{hermitian equation for kp}) in the case of $\bm{k} = \bm{K}_1$ as
\begin{eqnarray}
&&H_0 \ket{\Psi_K^{(i)}}=\omega_K^{(i)}\gamma\ket{\Psi_K^{(i)}}\ (i=1,2,3),
\label{hermitian equation for kp at K} \\
&&H_0(\bm{K}_1)=
\begin{pmatrix}
h_3(\bm{K}_1) &0 \\
0& I_3
\end{pmatrix}
, 
\end{eqnarray}
where $\ket{\Psi_K^{(i)}}=(\bm{U}^{(i)},\bm{V}^{(i)})^T$ with $ i = 1, 2, 3 $ corresponding to $\bm{A}_l^{(i)} $. Physically, $\bm{V}^{(i)}$ corresponds to the velocity. Equation~(\ref{hermitian equation for kp at K}) has already been solved in Sec.~\ref{sec:level2 subsec:level2}. 

\begin{figure}
\includegraphics[width=7.5cm]{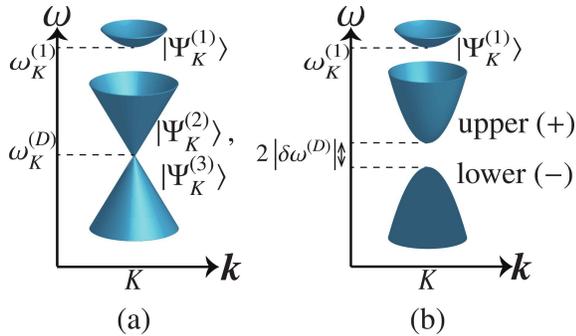}
\caption{\label{fig:dirac_cones} Schematic pictures of the band structures near the $K$ point, (a) when the time-reversal symmetry is preserved and (b) when the time-reversal symmetry is broken. In (b) the Dirac cone splits and has a gap $2|\delta \omega^{(D)}|$ at the $K$ point.}
\end{figure}

Here we calculate the dispersion relation slightly away from the $\bm{K}_1$ point, by setting $\mbox{\boldmath $k$} = \bm{K}_1 + \delta \mbox{\boldmath $k$} $ in the $ \mbox{\boldmath $k$}\cdot \mbox{\boldmath $p$} $ perturbation theory. When $\bm{k}$ is away from the $K$ point, the Hamiltonian deviates from $H_0$, and this deviation is written as
\begin{align}
\delta H&=
\begin{pmatrix}
\delta h_3 & 0 \\
0&0
\end{pmatrix}
, \\
(\delta h_3)_{ij}&=-(c_l^2 - c_t^2)i[ \delta k_{i}(\nabla_j+ iK_{1,j})+\delta k_{j}(\nabla_i+ iK_{1,i}) ]\nonumber \\
&\ \ \ \  -2ic_t^2\delta_{ij} \delta\mbox{\boldmath $k$}\cdot(\mbox{\boldmath $\nabla$} + i\bm{K}_1) 
\end{align}
by a straightforward calculation. Here, we study how the double degeneracy $\omega_K^{(D)}\equiv\omega_K^{(2)}=\omega_K^{(3)}$ at the $K$ point is lifted away from the $K$ point. Since the unperturbed eigenmodes 
$\ket{\Psi_K^{(2)}}$ and $\ket{\Psi_K^{(3)}}$
are degenerated at the $K$ point, we express the wavefunction as a linear combination $\ket{\Psi(\bm{k})}=a\ket{\Psi_K^{(2)}}+b\ket{\Psi_K^{(3)}}$ with coefficients $a$ and $b$, and we should solve
\begin{align}
\begin{pmatrix}
\braket{\Psi_K^{(2)}|\delta H|\Psi_K^{(2)}} & \braket{\Psi_K^{(2)}|\delta H|\Psi_K^{(3)}} \\
\braket{\Psi_K^{(3)}|\delta H|\Psi_K^{(2)}} & \braket{\Psi_K^{(3)}|\delta H|\Psi_K^{(3)}}
\end{pmatrix}
\begin{pmatrix}
a \\
b
\end{pmatrix}
=\delta \omega^{(D)}(\bm{k})N
\begin{pmatrix}
a \\
b
\end{pmatrix}, \label{perturbation matrix}
\end{align}
where $\delta \omega^{(D)}$ is a deviation of the eigenfrequency from $\omega_K^{(D)}$. By a direct calculation we obtain $\braket{\Psi_K^{(2)}|\delta H|\Psi_K^{(2)}}= \braket{\Psi_K^{(3)}|\delta H|\Psi_K^{(3)}}=0$, $\braket{\Psi_K^{(2)}|\delta H|\Psi_K^{(3)}}=(\braket{\Psi_K^{(3)}|\delta H|\Psi_K^{(2)}})^*=vN(-i\delta k_x + \delta k_y)$ and
\begin{eqnarray}
\delta \omega^{(D)}(\bm{k})=\pm v |\delta \mbox{\boldmath $k$}|,
\end{eqnarray}
where
\begin{widetext}
\begin{eqnarray}
v=\frac{3\sqrt{3}}{4N}a^2\left[ \frac{c_l^2+\beta^2 c_t^2}{\beta}+\frac{\beta_{\lambda}^2c_l^2+c_t^2}{\beta_{\lambda}^3} \nu^2 + \frac{((3 - \beta^2)\beta_{\lambda}+(1+\beta_{\lambda}^2)\beta)c_l^2+((3 - \beta_{\lambda}^2)\beta+(1+\beta^2)\beta_{\lambda})c_t^2}{\beta_{\lambda} (\beta+\beta_{\lambda})} \nu \right],
\label{v}
\end{eqnarray}
\end{widetext}
$N$ is given by Eq.~(\ref{N}) and $\omega$ in Eq.~(\ref{N}) denotes $\omega_K^{(D)}$ in Eq.~(\ref{omega_K^(D)}). This result shows that the degeneracy at the $K$ point is lifted when the wavevector is away from the $K$ point, and has a linear dispersion around the $K$ point. In other words, we have showed emergence of a Dirac cone at the $K$ point (Fig.~\ref{fig:dirac_cones} (a)). We can apply a similar method to show emergence of the Dirac cone also at the $K'$ point, which is naturally expected from sixfold rotational symmetry.
We note that in Ref.~\onlinecite{yu2016surface} a surface phononic graphene is experimentally realized, and Dirac cones are observed at
$K$ and $K'$ points. From the symmetry viewpoint, the emergence of the Dirac cone in our system is the same as the one in 
the surface phononic graphene, stemming from the threefold rotation and time-reversal symmetry, while the lattice structures are different.
\section{\label{sec:level3}TOPOLOGICAL BANDS OF SAWS BY BREAKING TIME-REVERSAL SYMMETRY}
We have seen that there appears double degeneracy at $K$ and $K'$ points. This degeneracy comes
from time-reversal and threefold rotational symmetries. Therefore, when time-reversal symmetry
is broken, this degeneracy can be lifted and the Dirac cones open a gap.
When Dirac cones split by adding some perturbations and open a gap, appearance of topological bands and topological edge modes within the gap are expected. 
This scenario has been seen in various systems in electronics \cite{Haldane}, photonics \cite{HaldaneRaghu}, phononics~\cite{PhysRevLett.114.114301,Mousavi,Wang-PRL,Wang-NJP,0022-3727-51-17-175302} and so on.

To show this possibility of realizing topological bands in the present system of SAWs, we introduce a term that breaks time-reversal symmetry into the equation of motion Eq.~(\ref{EOM}), and investigate behaviors of the Dirac cones at the $K$ and $K'$ points. Finally, we calculate the Chern number for each band to show topological nature of the bands and topological edge modes~\cite{PhysRevLett.49.405}.

In Appendices~\ref{gauge invariance of Berry curvature}, \ref{quantization of the Chern number} and \ref{bulk-edge correspondence}, we explain the details of basic topological properties of Chern number used in this section. We emphasize that these topological properties apply not only to discrete systems but also to continuum systems like the present case.
\subsection{\label{sec:level3 subsec:level1}Time-Reversal Symmetry Breaking due to the Coriolis Force}
In order to break time-reversal symmetry, we rotate the elastic body at a constant angular frequency $\Omega$ around the $z$ axis in the fixed inertial frame (see Fig.~\ref{fig:surface})~\cite{Wang-NJP}. Here, we assume that $\Omega$ is sufficiently small, and that the centrifugal force can be neglected, since the centrifugal force has a quadratic dependence on the angular frequency. Therefore, in the reference frame rotating together with the elastic body about the $z$ axis, we add only a term of the Coriolis force to Eq.~(\ref{EOM}) and get
\begin{eqnarray}
\ddot{\mbox{\boldmath $u$}}=c_t^2 \mbox{\boldmath $\nabla$}^2\mbox{\boldmath $u$}+(c_l^2 - c_t^2)\mbox{\boldmath $\nabla$}(\mbox{\boldmath $\nabla$}\cdot\mbox{\boldmath $u$})+2\dot{\mbox{\boldmath $u$}} \times \mbox{\boldmath $\Omega$}.
\end{eqnarray}
We note that the idea of breaking time-reversal symmetry by the Coriolis force has been adopted in Ref.~\onlinecite{Wang-NJP}.
In Sec.~\ref{sec:level2 subsec:level3}, we chose the normalization using the total energy. For that reason, if the system is changed, we need to redo the normalization. Eventually, we can use the same formula for the norm $N=\braket{\Psi_n(\bm{k})| \gamma |\Psi_n(\bm{k})}$ with the same matrix $\gamma$ as in Sec.~\ref{sec:level2 subsec:level3}, because the Coriolis force does not exert work on the system, like the Lorenz force in electronic systems. 
Therefore, we use
\begin{eqnarray}
&&(H_0+\delta V)\ket{\Psi^{(i)}_K}=(\omega^{(i)}_K + \delta \omega^{(i)}) \gamma \ket{\Psi^{(i)}_K}, \\
&&\delta V=
\begin{pmatrix}
2i\omega \Omega_3 &0 \\
0&0
\end{pmatrix}
,\ 
\Omega_3=
\begin{pmatrix}
0&\Omega&0 \\
-\Omega & 0 &0 \\
0&0&0
\end{pmatrix}
,
\end{eqnarray}
instead of Eq.~(\ref{hermitian equation for kp at K}). Here, $\delta V$ represents the term breaking time-reversal symmetry,  
and $\delta \omega^{(i)}$ denotes a corresponding deviation of the eigenfrequency from $\omega_K^{(i)}$. We investigate what happens to the degeneracy at the $ K $ point by introducing the perturbation $\delta V$. Therefore, we consider the cases with $i=2$ and $i=3$ and develop degenerate perturbation theory for the two eigenmodes. 
Calculations similar to Eq.~(\ref{perturbation matrix}) lead us to dispersion relations. By straightforward calculations in the lowest order in $\zeta$, we get
\begin{eqnarray}
\delta \omega^{(D)}=\pm \omega_G, \label{delta omegaD} 
\end{eqnarray}
where
\begin{eqnarray}
\omega_G=\frac{9\sqrt{3}a^2 d}{N} \omega_K^{(D)} \Omega \label{omegaG}
\end{eqnarray}
represents the gap due to breaking of the time-reversal symmetry. The details of derivation of Eq.~(\ref{omegaG}) are in Appendix~\ref{calculations of omegaG}. Thus, we have showed that the Dirac cones split by introducing the Coriolis force which breaks time-reversal symmetry (Fig.~\ref{fig:dirac_cones} (b)). 
There might also be other means to break the time-reversal symmetry, for example, by coupling with gyroscopes, which are spinning tops pinned to the lattice sites. It breaks the time-reversal symmetry by an inertial force.
\cite{Nash-PNAS,Wang-PRL}.
\subsection{\label{sec:level3 subsec:level3}Chern Number of the Bands of SAWs}
When the Dirac cones split by breaking the time-reversal symmetry, topological edge modes are expected to appear. In order to find out whether they appear, we calculate the Chern number. 
For this purpose, we 
find that in the present case, the Berry connection  $\bm{A}_n(\bm{k})$ and Berry curvature~\cite{Berry45} 
$\bm{B}_n(\bm{k})$ should be defined as 
\begin{align}
&\bm{A}_n(\bm{k}) \equiv i \braket{\Phi_n(\bm{k})|\gamma\bm{\nabla}_k  |\Phi_n(\bm{k})}
\\
&\bm{B}_n(\bm{k}) \equiv \bm{\nabla}_k \times \bm{A}_n(\bm{k}),
\end{align}
from the Bloch wavefunction $\ket{\Phi_n(\bm{k})}$, where $n$ denotes a band index. Here the wavefunction $\Phi_n(\bm{k})$ should
be normalized as $\langle\Phi_n(\bm{k})|\gamma|\Phi_n(\bm{k})\rangle=1$. 
Because our system is described by the generalized eigenvalue problem, the definition of the Berry connection and Berry curvature
is different from that in ordinary eigenvalue problems, and it contains an extra factor $\gamma$. With this factor, the Berry curvature possesses important properties such as gauge invariance,
similarly to the Berry curvature in an ordinary eigenvalue problem. It is explained in detail in Appendix~\ref{gauge invariance of Berry curvature}, where we see that the gauge invariance comes from the 
Hermiticity of the problem, i.e. the Hermiticity of the matrices $H$ and $\gamma$ in Eq.~(\ref{eigen eq}).

In two-dimensional systems, the Berry curvature has only the $z$ component $B_{n,z}(\bm{k}) \equiv \frac{\partial}{\partial k_x}A_{n,y}(\bm{k})
-\frac{\partial}{\partial k_y}A_{n,x}(\bm{k})$.
Then the Chern number for the $n$-th band is defined as an integral over the Brillouin zone:
\begin{eqnarray}
\mathcal{C}_n =\int_{\mathrm{BZ}} \frac{\mathrm{d} \bm{k}}{2\pi} B_{n,z}(\bm{k}). \label{Chern number}
\end{eqnarray}
The Chern number is quantized to be an integer, whenever the $n$-th band is separated from other bands by a gap. 
This quantization for generalized eigenvalue problems is shown in Appendix~\ref{quantization of the Chern number}, where we see that the Hermiticity of the problem plays an essential role.  
Nevertheless, so far we only know the wavefunctions near the $K$ point, which seems to be insufficient for calculations of Eq.~(\ref{Chern number}). Nonetheless, as we see in the following, we can calculate the Chern number when the gap $\omega_G$ is nonzero. 
For the calculation we note that when the time-reversal symmetry is preserved, the Berry curvature is zero everywhere,
because time-reversal symmetry gives $B_{n,z}(\bm{k})=-B_{n,z}(-\bm{k})$ and twofold rotational 
symmetry gives $B_{n,z}(\bm{k})=B_{n,z}(-\bm{k})$. When the time-reversal symmetry is slightly broken, the gap $\omega_G$ is small, and the Berry curvature is sharply concentrated around the $K$ and $K'$ points, 
as we see in the following. Therefore, the integral Eq.~(\ref{Chern number}) is well approximated by contributions around these points.

In order to calculate the Berry curvature, we need to calculate 
the eigenvectors away from the $K$ point without time-reversal symmetry. We develop degenerate perturbation theory with two perturbation terms, $\delta V$ and $\delta H$: 
We express the wavefunction $\Phi(\bm{k})$ at the wavevector away from $K$ as $\ket{\Phi(\bm{k})}=a\ket{\Psi_K^{(2)}}+b\ket{\Psi_K^{(3)}}$, and we get
\begin{align}
\begin{pmatrix}
-\omega_{\mathrm{G}}&v(-i\delta k_x+\delta k_y)\\
v(i\delta k_x+\delta k_y)&\omega_{\mathrm{G}}
\end{pmatrix}
\begin{pmatrix}
a \\b
\end{pmatrix}
= \delta\omega^{(D)} 
\begin{pmatrix}
a \\b
\end{pmatrix}
.
\end{align}
Since the eigenvalues of this equation are
\begin{eqnarray}
\delta \omega^{(D)}=\pm \sqrt{\left( v|\delta \mbox{\boldmath $k$} | \right)^2+\omega_G^2}\ ,
\end{eqnarray}
which have a gap $2|\omega_G|$ (Fig.~\ref{fig:dirac_cones}), in agreement with the results of the previous sections. The matrix in this equation has the same form as the massive Dirac Hamiltonian, and hence we get the Berry curvature
\begin{eqnarray}
B_{\pm,z}(\delta \mbox{\boldmath $k$})=\pm \frac{ v ^2\omega_{G}}{2\left( \left(v|\delta \mbox{\boldmath $k$}|\right)^2 + \omega_{G}^2  \right)^{3/2} }.
\label{eq:Berry}
\end{eqnarray}
Here, $B_{+,z}$ and $B_{-,z}$ represent the $z$ components of the Berry curvatures of the upper band and the lower band respectively. By using this we can calculate the Chern number. An integral of $B_{\pm,z}$ over a region near the $K$ point is equal to $\mathcal{C}_{\pm}(K)\simeq \pm(1/2){\rm sgn}(\omega_G)$. The contribution from the $K'$ point is identical with that from the $K$ point because of the sixfold rotational symmetry. Because 
the Berry curvature is sharply concentrated around the $K$ and $K'$ points, the resulting Chern number is
\begin{eqnarray}
\mathcal{C}_{\pm}=\mathcal{C}_{\pm}(K)+\mathcal{C}_{\pm}(K')\simeq\pm{\rm sgn}(\omega_G).
\label{eq:Chern}
\end{eqnarray}
While this is an approximated result by evaluating the integral only near the $K$ and $K'$ points, it is in fact exactly equal to $\pm{\rm sgn}(\omega_G)$ because $\mathcal{C}_{\pm}$ is quantized to be an integer. 
We explain more details of this calculation and related discussion in Appendix~\ref{quantization of the Chern number}. 

Suppose $\omega_G$ is positive, Then, with the term breaking time-reversal symmetry, the lowest band has the Chern number equal to $-1$, which means that there appears one branch of chiral edge modes within the gap, going along the edge in a counterclockwise way. 
This results from the bulk-edge correspondence. The bulk-edge correspondence is well established for Hermitian eigenvalue problems, and 
is shown for generalized Hermitian eigenvalue problems in Appendix~\ref{bulk-edge correspondence}. 
In Fig.~\ref{fig:surface} we show a schematic picture of the topological edge modes. These
topological edge modes appear along the edge of the corrugated surface. Irrespective of the 
detailed shape of this surface, the edge modes go along the edge of the system. 

We have ignored the centrifugal force in our theory. The centrifugal force gives rise to a expansion of
the system in the radial direction. It gives rise to a spatial variation of mass density, leading to a 
spatial variation of the frequency at  the Dirac point. 
Therefore, if this spatial variation is smaller than the gap size $\omega_G$, the gap is open  for the whole system, and 
one can safely ignore the
centrifugal force. This condition is discussed in detail in Appendix \ref{centrifugal}, and 
is obtaied as 
\begin{eqnarray}
\Omega \ll \frac{27\sqrt{3}a^2 dc_l^2}{N R^2}.
\end{eqnarray}
As an example, we take $R=1\mathrm{cm}$, $a=1\mathrm{mm}$, and $d=10\mathrm{\mu m}$. We put the velocities of the acoustic waves
to be $c_l=4.078\times10^4 \mathrm{m/s}$, $c_t=2.180\times10^4 \mathrm{m/s}$ from those of the material $\mathrm{Ni}$~\cite{yu2016surface}. Then this leads to a
condition to safely neglect the centrifugal force as $\Omega \ll 0.894 \times 10^{4} \mathrm{Hz}$. Thus if we put $\Omega=10^2\mathrm{Hz}$, the
gap size is $\omega_G=0.114\times10^2\mathrm{Hz}$ and the frequency of the Dirac point is $\omega_K^{(D)}=8.48\times 10^6\mathrm{Hz}$.

\begin{figure}
\includegraphics[width=8cm]{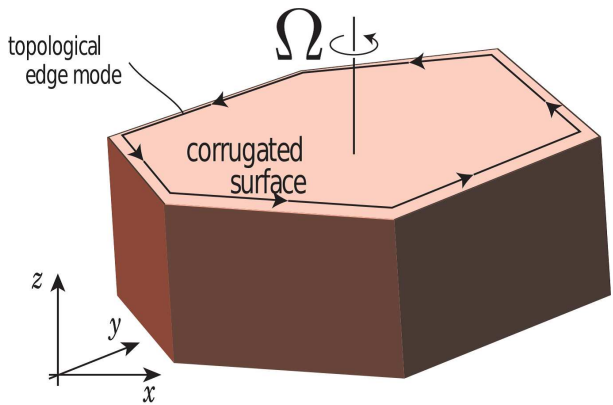}
\caption{\label{fig:surface}Schematic picture of the setup. The corrugated surface is within the $xy$ plane, and the 
system is rotated around the $z$ axis with the angular velocity $\Omega$. As a result of breaking of time-reversal symmetry, 
the Dirac cones at $K$ and $K'$ points will open a gap, and there will be topological edge modes going around the system in the counterclockwise way 
when the lowest bulk band has the Chern number ${\cal C}_{-}=-1$.}
\end{figure}
\section{\label{sec:level4}SUMMARY AND DISCUSSION}
In this paper we considered surface acoustic waves on a corrugated surface of an elastic body, forming a triangular lattice. Firstly, we calculated eigenmodes at the $K$ and $K'$ points by superposing eigenmodes for a flat surface at wavevectors which are mutually different by the reciprocal lattice vectors, and then calculated those around the $K$ and $K'$ points by the $\bm{k} \cdot \bm{p}$ perturbation theory. In the 
calculation, we rewrote the non-Hermitian eigenvalue equation into a generalized Hermitian eigenvalue equation by noting that the total energy is conserved. Eventually we showed emergence of Dirac cones at $K$ and $K'$ points. Then in order to open a gap we finally introduced a term which breaks time-reversal symmetry. Then the Dirac cones open a gap and the Chern number for the lowest band takes a non-zero value. Therefore, we showed that the elastic waves localized on the surface of an elastic body have topologically protected chiral edge modes. As a result, one-dimensional chiral elastic waves are realized along the edge of a surface of a three-dimensional elastic body (Fig.~\ref{fig:surface}). 

These proposals can be tested in simulations or in experiments. The important point of the present theory is wide applicability, with no strict 
restrictions on materials, frequencies and sizes of the unit cell. Moreover, in the present paper we introduced the Coriolis force to break the time-reversal symmetry, but there might be other means to break the time-reversal symmetry, for example, by coupling with gyroscopes
\cite{Nash-PNAS,Wang-PRL}.
 We emphasize that significance of the present paper lies not only in the resulting physical phenomena but also in its theoretical framework 
itself. We have established a theory of Berry curvature and topological bands for general acoustic waves including continuum systems, with an example of SAWs on a periodically corrugated surface. 
In this theory, the essential step is to transform the non-Hermitian eigenvalue problem into a generalized Hermitian eigenvalue problem, and this transformation can be performed by focusing on 
a conserved quantity, i.e. the energy. Furthermore, by treating the corrugation as a perturbation, we can analytically calculate the eigenmodes on the corrugated surface. This shows a microscopic mechanism
how the band structure for the periodic system (i.e. SAWs on a corrugated surface) is determined from that for a free space (i.e. SAWs on a flat surface). 
Thus our theory can be applied to any mechanical systems, including both discrete and continuum systems, and can serve as a building block for various kinds of topological band theory for 
acoustic waves. 

\begin{acknowledgments}
This work was supported by JSPS KAKENHI Grant Numbers JP26287062, and JP18H03678.
\end{acknowledgments}
\appendix
\section{\label{eigenvalues and eigenvectors of SAW}Calculations of the eigenvectors and eigenvalues of SAWs on the periodically corrugated surface}
In this appendix, we show the details of the calculations for eigenvectors and eigenvalues at the $K$ point in Sec.~\ref{sec:level2 subsec:level2}. 

As stated in Sec.~\ref{sec:level2 subsec:level2}, solutions of Eq.~(\ref{EOM}) in the region $z \leq \zeta$ are written as
\begin{widetext}
\begin{eqnarray}
&&u_x=\sum_{n_1,n_2}\frac{k_{n_1 n_2,x}}{k_{n_1n_2}}\left( A_{l,n_1 n_2} e^{\alpha_{l,n_1n_2} z} + A_{t,n_1 n_2} e^{\alpha_{t,n_1 n_2} z} \right) e^{i(\bm{k}_{n_1n_2}\cdot \bm{r}-\omega t)}\ , \label{u_x}\\
&&u_y=\sum_{n_1,n_2}\frac{k_{n_1 n_2,y}}{k_{n_1n_2}}\left( A_{l,n_1 n_2} e^{\alpha_{l,n_1n_2} z} + A_{t,n_1 n_2} e^{\alpha_{t,n_1 n_2} z} \right) e^{i(\bm{k}_{n_1n_2}\cdot \bm{r}-\omega t)}\ ,\label{u_y}\\
&&u_z=\sum_{n_1,n_2}\left( -\frac{\alpha_{l,n_1 n_2}}{k_{n_1 n_2}}A_{l,n_1 n_2} e^{\alpha_{l,n_1 n_2} z} - \frac{k_{n_1 n_2}}{\alpha_{t,n_1 n_2}}A_{t,n_1 n_2} e^{\alpha_{t,n_1 n_2} z} \right)i e^{i(\bm{k}_{n_1n_2}\cdot \bm{r}-\omega t)}\ ,
\label{u_z}
\end{eqnarray}
\end{widetext}
by hybridizing the plane-wave solutions on the flat surface with various wavevectors which are different from each other by the reciprocal lattice vectors of the surface, where $k_{n_1n_2}=|\mbox{\boldmath $k$}_{n_1n_2}|,\ \mbox{\boldmath $k$}_{n_1n_2}=\mbox{\boldmath $k$}+\mbox{\boldmath $G$}_{n_1n_2},\ \mbox{\boldmath $G$}_{n_1n_2}=n_1\mbox{\boldmath $b$}_1 + n_2 \mbox{\boldmath $b$}_2$ is a reciprocal lattice vector, $\alpha_{l,n_1n_2}=\sqrt{k^2_{n_1n_2} - \omega^2/c_{l}^2}$ and $\alpha_{t,n_1n_2}=\sqrt{k^2_{n_1n_2} - \omega^2/c_{t}^2}$. In the summation, $n_1$ and $n_2$ run over integers. $A_{l,n_1 n_2}$ and $A_{t,n_1 n_2}$ are constants which will be determined later. Henceforth, we write $n_1$ and $n_2$ together as $n$ for simplicity. 

To determine dispersions and the coefficients $A_{l,n}$ and $A_{t,n}$, we impose the stress-free boundary conditions $\sum_{j} \sigma_{ij} e_j |_{z = \zeta}=0\ (i = 1,2,3) $. We can write $\mbox{\boldmath $e$}$ as $\mbox{\boldmath $e$}=(-\zeta_x, -\zeta_y, 1) ((\zeta_x) ^ 2 + (\zeta_y) ^ 2 + 1)^{- 1/2}$, where $\zeta_{x}=\partial \zeta/\partial x$ and $\zeta_{y}=\partial \zeta/\partial y$. Hence, the boundary conditions are written as
\begin{eqnarray}
\left[ -\zeta_x \sigma_{xx} - \zeta_y \sigma_{xy} + \sigma_{xz}  \right]_{z=\zeta} &=&0\ , \label{BC2-1}\\
\left[ -\zeta_x \sigma_{yx} - \zeta_y \sigma_{yy} + \sigma_{yz}  \right]_{z=\zeta} &=&0\ , \label{BC2-2}\\
\left[ -\zeta_x \sigma_{zx} - \zeta_y \sigma_{zy} + \sigma_{zz}  \right]_{z=\zeta} &=&0\ . \label{BC2-3}
\end{eqnarray}
By writing $\sigma_{ij}$ in terms of the displacement vector $\bm{u}$ given by Eqs.~(\ref{u_x})-(\ref{u_z}), we obtain the dispersion relation in the following. In the calculation, we use  Fourier expansion in the $xy$ plane with $\bm{r}=(x,y),$
\begin{eqnarray}
&&e^{\alpha_{n} \zeta}=\sum_m C_{\bm{G}_m}(\alpha_{n})e^{i\bm{G}_m \cdot \bm{r}}, \\
&&C_{\bm{G}_m}(\alpha_{n})=\frac{1}{S}\int_{\mathrm{unit\ cell}} \mathrm{d}^2\bm{r}\ e^{\alpha_{n} \zeta}e^{-i\bm{G}_m \cdot \bm{r}}, \label{fourier coefficient}
\end{eqnarray}
and its derivatives with respect to $x$ and $y$:
\begin{eqnarray}
&&\zeta_x e^{\alpha_{n} \zeta}=\sum_m i \frac{G_{m,x}}{\alpha_{n}} C_{\bm{G}_m} e^{i \bm{G}_m \cdot \bm{r}} , \\
&&\zeta_y e^{\alpha_{n} \zeta}=\sum_m i \frac{G_{m,y}}{\alpha_{n}} C_{\bm{G}_m} e^{i \bm{G}_m \cdot \bm{r}}.
\end{eqnarray}
Here, $\bm{G}_m=m_1 \bm{b}_1 + m_2 \bm{b}_2$ and $m=(m_1,m_2)$, the integration in Eq.~(\ref{fourier coefficient}) is performed in the surface unit cell, $S$ denotes an area of the surface unit cell and $\alpha_{n}$ denotes $\alpha_{l,n}$ or $\alpha_{t,n}$. Next we define integers $p_i=n_i+m_i\ (i=1,2)$, and replace $\sum_m$ with $\sum_p$. Then, Eqs.~(\ref{BC2-1})-(\ref{BC2-3}) are rewritten as 
\begin{widetext}
\begin{eqnarray}
&&0=\sum_{n}\frac{1}{k_n} \left( \frac{1}{\alpha_{l,n}} \left[ \omega^2 \left(k_{p,x}-k_{n,x} \right)+2c_t^2\left( \alpha_{l,n}^2 k_{p,x} +k_{n,y}(\bm{k}_n \times \bm{k}_p)_z \right) \right] C_{\bm{G}_{p-n}}(\alpha_{l,n}) A_{l,n} \right. \label{BC1} \nonumber \\
&&\left.\ \ \ \ \ \ +\frac{k_{n,x}}{\alpha_{t,n}}\left[ -\omega^2 +2c_t^2 \bm{k}_{n} \cdot \bm{k}_{p} \right] C_{\bm{G}_{p-n}}(\alpha_{t,n}) A_{t,n}  \right), \\
&&0=\sum_{n}\frac{1}{k_n} \left( \frac{1}{\alpha_{l,n}} \left[ \omega^2 \left(k_{p,y}-k_{n,y} \right)+2c_t^2\left( \alpha_{l,n}^2 k_{p,y} +k_{n,x}(-\bm{k}_n \times \bm{k}_p)_z \right) \right] C_{\bm{G}_{p-n}}(\alpha_{l,n}) A_{l,n} \right. \label{BC2} \nonumber \\
&&\left.\ \ \ \ \ \ +\frac{k_{n,y}}{\alpha_{t,n}}\left[ -\omega^2 +2c_t^2 \bm{k}_{n} \cdot \bm{k}_{p} \right] C_{\bm{G}_{p-n}}(\alpha_{t,n}) A_{t,n}  \right), 
\\
&&0=\sum_{n} \frac{1}{k_n} \left( \left[ \omega^2 - 2c_t^2 \bm{k}_{n} \cdot \bm{k}_{p} \right] C_{\bm{G}_{p-n}}(\alpha_{l,n}) A_{l,n} + \frac{1}{\alpha_{t,n}^2} \left[ k_n^2 \omega^2 - c_t^2 \bm{k}_{n} \cdot \bm{k}_{p} \left( \alpha_{t,n}^2+k_n^2 \right) \right] C_{\bm{G}_{p-n}} (\alpha_{t,n}) A_{t,n}  \right), \label{BC3}
\end{eqnarray}
\end{widetext}
which are to be satisfied for all integers $p_j=0,\pm1,\pm2,...\ .$ 

Equations~(\ref{BC1})-(\ref{BC3}) can be regarded as infinite-dimensional matrix equations. In order to solve them analytically, we focus on the $K$ point in the Brillouin zone. At the $K$ point, when we perturbatively switch on the surface corrugation, the plane wave at $\bm{K}_1=4\pi/(3a)(0,1)$ mixes with those at other $K$ points $\bm{K}_2=4\pi/(3a)(-\sqrt{3}/2,-1/2)$ and $\bm{K}_3=4\pi/(3a)(\sqrt{3}/2,-1/2)$ (Fig.~\ref{fig:surface_and_Kpoint}). In the lowest order in the perturbation theory, we have to consider the terms only from $\bm{K}_1$, $\bm{K}_2$ and $\bm{K}_3$, and we can ignore contribution from other $\bm{k}$ points in the summations in Eqs.~(\ref{BC1})-(\ref{BC3}). In addition, we approximate the ratio of the longitudinal amplitude to the transverse amplitude to be the value at the zeroth order of the perturbation because $\zeta$ is small~\cite{glass}. In other words, we approximate the ratio to be equal to that for a flat surface. This approximation indicates that $A_{t,n}\simeq (-1+\xi^2/2)A_{l,n}\equiv \nu A_{l,n}$. Furthermore, because the corrugation is treated as a perturbation, we can approximate Eq.~(\ref{fourier coefficient}) as
\begin{eqnarray}
C_{\bm{G}_m}(\alpha_{n})=\delta_{m,0} + \frac{\alpha_{n}}{S} \int \mathrm{d}^2 \mbox{\boldmath $r$}\ \zeta(\mbox{\boldmath $r$}) e^{-i\bm{G}_m \cdot \bm{r}}. \label{fourier coefficient 2}
\end{eqnarray}
Here, we can safely set $C_{\bm{0}}=1$, because the second term of Eq.~(\ref{fourier coefficient 2}) represents the average of the height of the surface, and it can be set to zero since physics is not affected by it. Moreover from Eq.~(\ref{fourier coefficient 2}) we obtain $ C_{\bm{b}_1}(\alpha)=C_{\bm{b}_2}(\alpha)=\alpha d/2$. By using these approximations for Eq.~(\ref{BC3}), we can obtain the eigenvectors
\begin{eqnarray}
&&\bm{A}_l^{(1)}=(1,\ 1,\ 1) \ , \label{A_l^1 in appendix}\\
&&\bm{A}_l^{(2)}=(1,\ \eta,\ \eta^{2}) \ , \label{A_l^2 in appendix}\\
&&\bm{A}_l^{(3)}=(1,\ \eta^{2},\ \eta) \ , \label{A_l^3 in appendix}
\end{eqnarray}
where $\eta=e^{2\pi i/3}$, and the eigenvalues 
\begin{eqnarray}
\omega_K^{(1)}&=&\sqrt{\frac{\xi^4 + 2d\beta (\beta \beta_{\lambda} - \nu^2)K}{\xi^2 -2d\beta(\beta \beta_{\lambda} + \nu)K}}c_t K, \\
\omega_K^{(2)}=\omega_K^{(3)} &=& \sqrt{\frac{\xi^4 -d\beta (\beta \beta_{\lambda} - \nu^2)K}{\xi^2 +d\beta(\beta \beta_{\lambda} + \nu)K}}c_t K \label{omega_K^(D) in appendix}
\end{eqnarray}
at the $K$ point, where $K=4\pi/(3a),\ \beta=\sqrt{1-\xi^2},\ \beta_{\lambda}=\sqrt{1-\xi^2\lambda^2}$, $\nu=-1+\xi^2/2$ and $\lambda=c_t/c_l$. The components of $\bm{A}_l^{(n)}$, $A_{l,i}^{(n)}\ (i=1,2,3)$, correspond to the $K_i$ points and the superscripts $(n)\ (n=1,2,3)$ are labels for the eigenmodes. Since the eigenmodes with $n = 2$ and $n=3$ are degenerate and form a doublet, we write $\omega_K^{(D)} \equiv \omega_K^{(2)}=\omega_K^{(3)}$. Thus, the triply degenerate modes at the $K$ point on the flat surface are lifted to a singlet ($\omega_K^{(1)}$) and a doublet ($\omega_K^{(2)}$ and $\omega_K^{(3)}$).

\section{\label{calculations of omegaG} Calculations of the band gap by the Coriolis force}
In this appendix, we calculate the gap by the Coriolis force, in the lowest order in the rotation frequency $\Omega$ and in the corrugation $d$. The eigenvalue equation with the Coriolis force is given by 
\begin{eqnarray}
&&(H_0+\delta V)\ket{\Psi^{(i)}_K}=(\omega^{(i)}_K + \delta \omega^{(i)}) \gamma \ket{\Psi^{(i)}_K}, \\
&&\delta V=
\begin{pmatrix}
2i\omega \Omega_3 &0 \\
0&0
\end{pmatrix}
,\ 
\Omega_3=
\begin{pmatrix}
0&\Omega&0 \\
-\Omega & 0 &0 \\
0&0&0
\end{pmatrix}
.
\end{eqnarray}
Here, $\delta V$ represents the term breaking time-reversal symmetry,  
and $\delta \omega^{(i)}$ denotes a corresponding deviation of the eigenfrequency from $\omega_K^{(i)}$. We investigate what happens to the degeneracy at the $ K $ point by introducing the perturbation $\delta V$. Therefore, we consider the cases with $i=2$ and $i=3$ and develop degenerate perturbation theory for the two eigenmodes. 
Calculations similar to Eq.~(\ref{perturbation matrix}) lead us to dispersion relations. 
To get the band gap $\omega_G$ in Eq.~(\ref{omegaG}), we calculate $\braket{\Psi_K^{(i)}|\delta V|\Psi_K^{(j)}} = \int_{\mathrm{unit\ cell}} dV \left(\Psi_K^{(i)} \right)^{\dag} \delta V \Psi_K^{(j)}$ $(i,j=2,3)$. Here, the integral in the unit cell is given by $\int _{\mathrm{unit\ cell}} dV=\int dS \int_{-\infty}^{\zeta(\bm{r})} dz$, where $\int dS$ denotes the integration over the unit cell in $xy$ plane. In the calculation, we first note that in the zeroth order in the corrugation $d$, this integral vanishes: 
$\int dS \int_{-\infty}^{0} dz\left( \Psi_K^{(i)} \right)^{\dag} \delta V \Psi_K^{(j)}=0$ for all values of $i$ and $j$. In the first order in $d$, the upper end of the integral over $z$ becomes $\zeta(\bm{r})$, leading to a nonzero result. 
Then, we get $\braket{\Psi_K^{(3)}|\delta V|\Psi_K^{(3)}}=-\braket{\Psi_K^{(2)}|\delta V|\Psi_K^{(2)}}=\omega_G N$,\ $\braket{\Psi_K^{(2)}|\delta V|\Psi_K^{(3)}}=\braket{\Psi_K^{(3)}|\delta V|\Psi_K^{(2)}}=0$ and 
\begin{eqnarray}
\omega_G=\frac{9\sqrt{3}a^2 d}{N} \omega_K^{(D)} \Omega,
\end{eqnarray}
which is identical with Eq.~(\ref{omegaG}). Thus the gap appears in the order $d^1 \Omega^1$.

\section{\label{gauge invariance of Berry curvature}Berry curvature for generalized Hermitian eigenvalue problems}
In the main text, we transformed the non-Hermitian eigenvalue problem $\tilde{H}\psi=\omega\psi$ 
into
the generalized Hermitian eigenvalue problem (\ref{eigen eq}), i.e. $H\psi=\omega\gamma\psi$. In the Bloch form it is rewritten as Eq.~(\ref{hermitian equation for kp}), i.e. 
\begin{equation}
H_0\Psi=\omega\gamma \Psi.
\label{eq:gen-eig}
\end{equation} Here $H_0$ and $\gamma$ are both Hermitian. 
One of the purposes of this transformation into a generalized Hermitian eigenvalue problem is that it is 
convenient for formulating
the Berry curvature.
In order to define the Berry curvature in non-Hermitian eigenvalue problems,
one needs to introduce two types of eigenvectors, i.e. left and right eigenvectors as was done in Ref.~\onlinecite{PhysRevLett.105.225901}, whereas
in Hermitian eigenvalue problems there is no need to distinguish between left and right eigenvectors.

Nonetheless, the other, and more important purpose of this transformation from a non-Hermitian eigenvalue problem into a generalized 
Hermitian eigenvalue problem is to ensure that the Berry curvature is well-defined, and that it possesses important necessary properties as Berry curvature, vital for physical phenomena as we explain in the following. 

One of the important properties of the Berry curvature is gauge invariance, which 
guarantees that the Berry curvature is an observable. 
Because the wavefunctions are normalized as $N=\braket{\Psi_n(\bm{k})| \gamma |\Psi_n(\bm{k})}$, the eigenvector for Eq.~(\ref{eq:gen-eig}) can have a gauge degree of freedom 
$\Psi_n(\bm{k}) \rightarrow \Psi'_n(\bm{k})\equiv e^{i\theta(\bm{k})}\Psi_n(\bm{k})$, where $\theta(\bm{k})$ is
an arbitrary real function of the wavevector $\bm{k}$. 
Then one can easily show that the Berry connection and the Berry curvature are transformed as
\begin{align}
\bm{A}'_n&=
 i \braket{\Psi'_n(\bm{k})|\gamma\bm{\nabla}_k  |\Psi'_n(\bm{k})}\nonumber\\
&= i \braket{\Psi'_n(\bm{k})|\gamma\bm{\nabla}_k  |\Psi'_n(\bm{k})}-\bm{\nabla}\theta(\bm{k})\nonumber\\
&=\bm{A}_n-\nabla_{\bm{k}}\theta,
\end{align}
and 
$B'_{n,z}(\bm{k})=B_{n,z}(\bm{k})$, Thus the Berry curvature is invariant under gauge transformation; it qualifies the Berry curvature to be an observable. 

It is known that the Berry curvature affects dynamics of a wavepacket through
the semiclassical equation of motion \cite{Sundaram,Okamoto}
\begin{align}
\dot{\bm{r}}=-\dot{\bm{k}}\times
\bm{\bm B}_{n}(\bm{k})
+\frac{\partial \omega_{n}}{\partial \bm{k}}.
\label{eq:eom}\end{align}
In the present case of two-dimensional systems, ${\bm B}_{n}(\bm{k})=(0,0,B_{n,z}(\bm{k}))$.
This semiclassical equation of motion has been formulated and widely used for Hermitian eigenvalue problems \cite{Sundaram}. 
In Ref.~\onlinecite{Okamoto}, this equation of motion is shown to be applicable also to generalized Hermitian eigenvalue 
problems. In its proof, conservation of the norm  $N=\braket{\Psi_n(\bm{k})| \gamma |\Psi_n(\bm{k})}$
coming from the Hermiticity of the problem is essential
\cite{Okamoto}.
Physically, the Hermiticity guarantees that in time evolution, the wavepacket never 
disappears and it is meaningful to consider its equation of motion.

In addition, in showing existence of topological edge modes in systems with a nonzero Chern number as explained in Appendix~\ref{bulk-edge correspondence},
the Hermiticity of the problem is essential. 
Thus, in various physical phenomena governed by the Berry curvature, the Hermiticity of the eigenvalue problem
is
required. 

\section{\label{quantization of the Chern number}Quantization of the Chern number for generalized Hermitian eigenvalue problems}
As discussed in the main text, when the $n$-th band is separated from other bands with a nonzero gap, the Chern number for the $n$-th band 
\begin{equation}
{\cal C}_n\equiv\int_{\mathrm{BZ}}\frac{\mathrm{d}\bm{k}}{2\pi}B_{n,z}(\bm{k})
\end{equation}
is quantized as an integer \cite{PhysRevLett.49.405,Kohmoto85}, and it represents the number of branches of chiral edge modes along the edge of the system in a clockwise way \cite{PhysRevB.23.5632}, as explained in detail in Appendix~\ref{bulk-edge correspondence}.
This quantization has been well studied in the context of integer quantum Hall systems of electrons, where the
governing equation is the Schr\"{o}dinger equation, i.e. a Hermitian eigenvalue equation.
In this paper, we are considering the generalized Hermitian eigenvalue problem (\ref{eq:gen-eig}), not
a simple Hermitian eigenvalue problem. Therefore, in this appendix, we show the quantization of the Chern number for generalized Hermitian eigenvalue problems, which 
is a generalization of the proof for usual Hermitian eigenvalue problems. In fact it was shown in the context of optics \cite{HaldaneRaghu}, and
here we show this quantization in generalized eigenvalue problems described by Eq.~(\ref{eq:gen-eig}), which can also apply to optics.

\begin{figure}
\includegraphics[width=8cm]{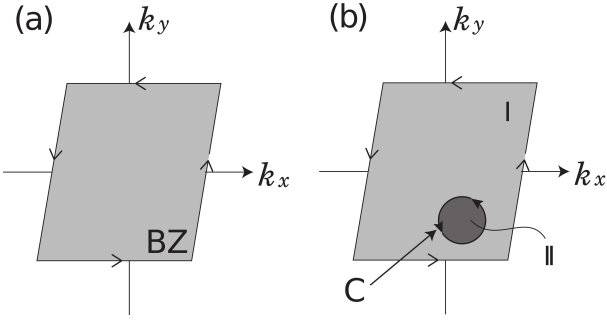}
\caption{\label{fig:Chern} Calculation of the Chern number as an integral over the Brillouin zone. (a) When a single gauge is chosen for the entire Brillouin zone, 
the integral of the Berry curvature is rewritten as a contour integral along the boundary of the Brillouin zone, leading to ${\cal C}_n=0$. (b) In some systems, one should divide the Brillouin zone into two regions I and II, in each of which the gauge is defined smoothly.}
\end{figure}

Here we show the quantization of the Chern number. 
First, if one can take a gauge which covers the entire Brillouin zone, i.e. if one can 
choose a wavefunction $\Psi_n(\bm{k})$ which is smooth over the entire Brillouin zone, 
we can use the Stokes' theorem to 
show that
\begin{equation}
{\cal C}_n\equiv\int_{\mathrm{BZ}}\frac{\mathrm{d}\bm{k}}{2\pi}B_{n,z}(\bm{k})
=\oint_{\partial(\mathrm{BZ})}\frac{\mathrm{d}\bm{k}}{2\pi}\cdot\bm{A}_n(\bm{k}).
\end{equation}
Here $\mathrm{BZ}$ denotes the two-dimensional Brillouin zone, and $\partial(\mathrm{BZ})$ does its one-dimensional boundary.
The above formula is  a contour integral along the boundary of the Brillouin zone (see Fig.~\ref{fig:Chern}(a)). Thanks to the 
Brillouin zone periodicity, this vanishes. On the other hand, in some systems, one cannot choose a
single gauge which covers the whole Brillouin zone with preserving the Brillouin zone periodicity.
Then the Brillouin zone should be divided into to regions I and II, in each of which the gauge is defined smoothly (see Fig.~\ref{fig:Chern}(b)). Let $u_n^{\rm I}(\bm{k})$ and $u_n^{\rm II}(\bm{k})$ denote the wavefunctions
 in the regions I and II, respectively, and $\bm{A}^{\rm I}(\bm{k})$ and $\bm{A}^{\rm II}(\bm{k})$
denote the corresponding Berry connection. Then, the Stokes' theorem leads to the following result
\begin{align}
{\cal C}_n&\equiv\int_{\mathrm{BZ}}\frac{\mathrm{d}\bm{k}}{2\pi}B_{n,z}(\bm{k})
=\sum_{i={\rm I},{\rm II}}\int_{i}\frac{\mathrm{d}\bm{k}}{2\pi}B_{n,z}(\bm{k})\nonumber\\
&=\oint_{\rm C}\frac{\mathrm{d}\bm{k}}{2\pi}\cdot(\bm{A}_{n}^{\rm II}(\bm{k})-\bm{A}_{n}^{\rm I}(\bm{k})),
\end{align}
where C is the loop forming the boundary between the two regions I and II. On the loop C, the wavefunctions $\Phi_n^{\rm I}(\bm{k})$ and $\Phi_n^{\rm II}(\bm{k})$
are different by a phase, $\Phi_n^{\rm I}(\bm{k})=e^{i\theta_n(\bm{k})}\Phi_n^{\rm II}(\bm{k})$, where 
$\theta_n(\bm{k})$ is real. Then it yields $\bm{A}_{n}^{\rm II}(\bm{k})-\bm{A}_{n}^{\rm I}(\bm{k})=\frac{\partial
\theta_n}{\partial \bm{k}}$, and 
\begin{equation}
{\cal C}_n=\oint_{\rm C}\frac{\mathrm{d}\bm{k}}{2\pi}\cdot\frac{\partial
\theta_n}{\partial \bm{k}}=\frac{1}{2\pi}\left[\theta_n\right]_{\rm C},
\end{equation}
where $\left[\theta_n\right]_{\rm C}$ represents a change of $\theta_n$ in going around the loop C. Because of the
single-valuedness of $\Phi_n^{i}$ ($i={\rm I},{\rm II})$ 
in the respective regions, this change of $\theta_n$ is an integer multiple of 
$2\pi$. Thus we conclude that the Chern number is quantized as an integer for generalized Hermitian eigenvalue problems. We note that the Hermiticity guarantees 
the gauge invariance of the Berry curvature, which in turn means gauge invariance of the Chern number. 

Next we introduce a parameter $m$ into the Hamiltonian $H(\bm{k})$, and 
suppose we change $m$ continuously. We focus on the Chern number ${\cal C}_n$ of the $n$-th 
band, and consider a change of the value of ${\cal C}_n$ as we change the parameter $m$. 
We assume that the $n$-th band is separated from other bands by a gap; then the quantization of
the Chern number means that the Chern number cannot change continously, and we conclude that the 
Chern number remains constant.
This is a fundamental property of the Chern number as a topological number. 

This property of the Chern number upon a change of the parameter $m$ helps us to calculate the Chern number analytically in some cases, without evaluating the Chern number as an integral over the Brillouin zone. In the present system of SAWs, we 
regard the frequency $\omega_{G}$, representing the breaking of time-reversal symmetry, 
as the external parameter $m$. Then as long as $\omega_G$ remains
positive, the gap is open, and the Chern number for the lowest band remains constant. Therefore, to 
evaluate the Chern number for any positive values of $\omega_G$, 
we can set $\omega_{G}$ to be a very small positive value to evaluate the Chern number. In that case, the distribution 
of the Berry curvature (\ref{eq:Berry}) is concentrated within the small region $|\delta\bm{k}|<|\omega_G|/v$ around the
$K$ and $K'$ points, and
the integral of the Berry curvature is well approximated by the integral around the $K$ and $K'$ points. 
As $\omega_G$ becomes smaller, the distribution of the Berry curvature (\ref{eq:Berry}) becomes sharper and its peak becomes larger, with its integral remains constant. 
Because the Berry curvature for $\omega_G=0$ is zero everywhere because of the sixfold rotation and time-reversal symmetries,
the result for the Chern number (\ref{eq:Chern}) asymptotically becomes exact when $\omega_G$ 
approaches positive infinitesimal. Thus, when $\omega_G$ is small the Chern number for the lowest band is $-1$, and
it means that the Chern number remains $-1$ even when $\omega_G$ takes any positive value.

\section{\label{bulk-edge correspondence}Bulk-edge correspondence for generalized Hermitian eigenvalue problems}
It is well established that the Chern number represents a number of branches of chiral edge modess along the edge of the system in a clockwise way 
\cite{PhysRevB.23.5632}. This is a seminal result known as bulk-edge correspondence,  
in the context of the integer quantum Hall effect in electronic systems,
and can be explained in various ways. This result is not limited to 
electronic systems but it holds also for any systems described by Hermitian eigenvalue equations, 
as has been applied for systems with various particles and quasiparticles \cite{Shindou1,Shindou2,PhysRevLett.105.225901,
Wang,HaldaneRaghu}.

\begin{figure}
\includegraphics[width=9cm]{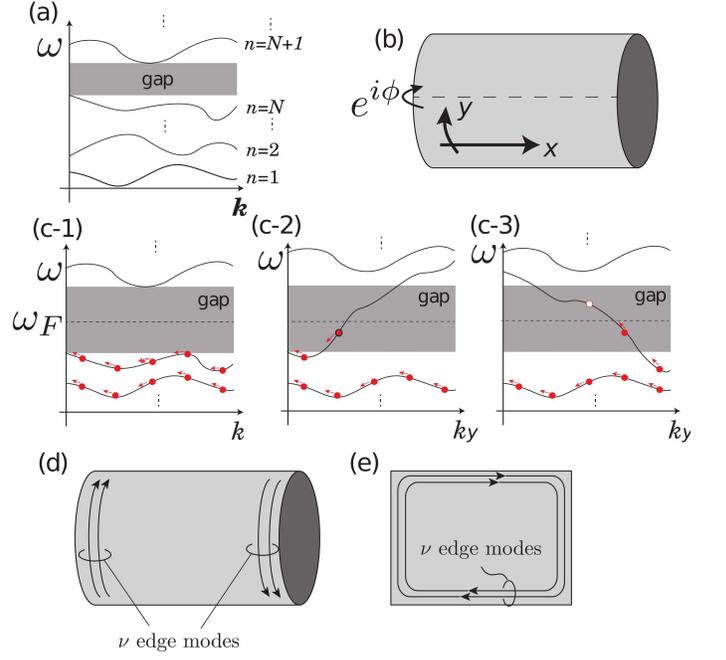}
\caption{\label{fig:Laughlin} Setup of the Laughlin's gedanken experiment. (a) Bulk band structure. (b) Schematic picture representing the
boundary condition imposed in the system.  (c-1)-(c-3) Flows of the modes under the change of the phase 
twist $\phi$ in the boundary conditions. The panels represent the cases (c-1) in the absence of modes in the gap,
(c-2) in the presence of a branch  with positive dispersion, and (c-3) in the presence of a branch with negative dispersion.
In (c-1)-(c-3), the red dots represent the modes allowed by the boundary condition below $\omega_F$, and we put one particle per eigenmode 
below $\omega_F$, i.e. per red dot. By changing $\phi$ the allowed modes
flow along the bands towards the directions specified by red arrows. After one cycle of $\phi$ from $0$ to $2\pi$, the
set of the red dots returns to the original one, except for (c-2) where the number of allowed modes decreases by one, 
and for (c-3) where it increases by one. (d) Edge modes in the cylinder geometry, whose existence is shown from the Laughlin gedanken experiment. 
(e) Edge mode in the open geometry.}
\end{figure}

To be precise, when the band structure has a gap, the sum $\nu$ of the Chern numbers of the bands below the gap is 
equal to the number of branches of edge modes inside the gap, which goes around the system in a clockwise way (see Fig.~\ref{fig:Laughlin}(e)). 
Suppose we consider a band structure in Fig.~\ref{fig:Laughlin}(a) and focus on the band gap between 
the $N$-th and $(N+1)$th bands, $\nu$ is given by $\nu=\sum_{n=1}^{N}{\cal C}_{n}$.
If $\nu$ is a negative integer, $|\nu|=-\nu$ represents the number of branches of edge modes in  a counterclockwise way. 
Because the Chern number is determined by the bulk wavefunctions, this correspondence 
is called bulk-edge correspondence, as can be shown by the Laughlin's gedanken experiment \cite{PhysRevB.23.5632}.
It is originally shown in the context of integer quantum Hall effect in electronic systems. Nonetheless,
it is not limited to electronic systems, but is common for various particle systems
described by Hermitian eigenvalue equations. 

Following the idea of the Laughlin's gedanken experiment, here we
show bulk-edge correspondence  for generalized Hermitian eigenvalue problems, including SAWs in this paper. 
Consider a two-dimensional system 
along the $xy$ plane, described by the generalized Hermitian eigenvalue problem Eq.~(\ref{eq:gen-eig}). 
Along the $x$ direction, we set an open boundary condition, and along the $y$ direction we
set 
a boundary condition with a $e^{i\phi}$ ($\phi$: constant) phase change when we go across the boundary in the
$+y$ direction. 
In particular, $\phi=0$ represents a periodic boundary condition. 
Thus the system can be thought of as an open cylinder (Fig.~\ref{fig:Laughlin}(b)), where wavefunctions change by 
a phase $e^{i\phi}$ across the broken line. 
In this setup, we consider a change in the position of the particle upon an adiabatic change of the phase $\phi$ from $0$ to $2\pi$. 
In electronic systems, it can be conveniently described in terms of a polarization, using the so-called
modern theory of polarization \cite{Resta,KingSmith,Vanderbilt}. 

As is similar to electronic systems  \cite{Resta,KingSmith,Vanderbilt}, 
when we consider this system as a one-dimensional system, one can construct a Wannier orbital which is 
localized around $x=R$ in the $x$ direction, 
where $R$ is a lattice translation vector. 
From the Wannier orbital, one can calculate an
expectation value of the center of the particle distribution as
\begin{equation}
x_n=\frac{a_x}{2\pi}\int_0^{2\pi/a_x}\mathrm{d}k_x\langle\Phi_{n,k_x}|i\gamma\frac{\partial}{\partial k_x}|\Phi_{n,k_x}\rangle,
\label{eq:xn}\end{equation}
where  $a_x$ being the unit cell size along the $x$ direction.
We here note that in the present system of SAWs, the center of the distribution is defined with respect to the energy density, 
which leads to the extra factor $\gamma$ in Eq.~(\ref{eq:xn}).
The normalization is taken as $\langle\Phi_{n,k_x}|\gamma|\Phi_{n,k_x}\rangle=1$.

Suppose we consider band structure with a gap, and let $N$ denote the number of bands 
below the gap considered. We fix one arbitray value of a frequency $\omega_{\rm F}$ inside the
gap as a reference frequency, and then put one particle per each mode below this reference frequency $\omega_{\rm F}$.
For fermions it is naturally realized by setting the Fermi energy inside the gap and setting the temperature to be
zero. For bosons such as phonons, it is an artificial procedure only for a proof of existence of topological edge modes. 
Then, the sum of the positions of the particles divided by the volume is 
\begin{equation}
P_x=\sum_{n=1}^{N}\frac{1}{2\pi}\int_0^{2\pi}\mathrm{d}k_x 
\langle\Phi_{n,k_x}|i\gamma\frac{\partial}{\partial k_x}|\Phi_{n,k_x}\rangle.
\end{equation}
Note that for fermions it is equal to the polarization divided by the electronic charge.

Let us change the phase $\phi$ describing the boundary condition. Then the perturbation theory tells us how the 
``polarization'' $P_x$ changes.
Due to the translational symmetry along $k_y$, the Bloch wavefunctions $\Phi_{n,k_x}$ are labeled
also by the wavenumber along $k_y$, and are written as $|\Phi_{n,\bm{k}}\rangle$ with $\bm{k}=(k_x,k_y)$. Let 
$L_y$ denote the size of the system along the $y$ direction. At $\phi=0$, $k_y$ is quantized as 
$k_y=2\pi j/L_y$ ($j$: integer). Then the phase twist $\phi$ shifts the quantized values of $k_y$
to 
\begin{equation}
k_y=\frac{2\pi j-\phi}{L_y} \ \ (j:\ \textrm{integer}).
\label{eq:ky}
\end{equation}
Then the change of the ``polarization'' $P_x$ upon the change of $\phi$ is given by
\begin{align}
&\delta P_x
=i\delta\phi \sum_{n=1}^{N}\int_0^{2\pi/a_x}\frac{\mathrm{d}k_x}{2\pi}\sum_{j}\nonumber\\
&
\left[
\left\langle\frac{\partial \Phi_{n,\bm{k}}}{\partial \phi}\right|\gamma\left|\frac{\partial \Phi_{n,\bm{k}}}{\partial k_x}\right\rangle
-\left\langle\frac{\partial \Phi_{n,\bm{k}}}{\partial k_x}\right|\gamma\left|\frac{\partial \Phi_{n,\bm{k}}}{\partial \phi}\right\rangle
\right]\end{align}
where $k_y$ is given by Eq.~(\ref{eq:ky}).
By integrating with respect to $\phi$, we get a net change of the ``polarization''
\begin{align}
\nu&\equiv \Delta P_x
=i\sum_{n=1}^{N}\int_{\mathrm{BZ}}\frac{\mathrm{d}\bm{k}}{2\pi}
\nonumber\\
&\ \ \ \ \ \ \left[
\left\langle\frac{\partial \Phi_{n,\bm{k}}}{\partial k_y}\right|\gamma\left|\frac{\partial \Phi_{n,\bm{k}}}{\partial k_x}\right\rangle
-\left\langle\frac{\partial \Phi_{n,\bm{k}}}{\partial k_x}\right|\gamma\left|\frac{\partial \Phi_{n,\bm{k}}}{\partial k_y}\right\rangle
\right]\nonumber\\
&=\sum_{n=1}^{N}\int_{\mathrm{BZ}}\frac{\mathrm{d}\bm{k}}{2\pi}B_{n,z}(\bm{k})=\sum_{n=1}^{N}{\cal C}_n,
\end{align}
where BZ denotes the two-dimensional Brillouin zone. 
It is an integer because the Chern number is an integer. Namely, in this change of $\phi$, $\nu$ particles are transferred along the $x$ direction.

Let us consider a case with $\nu$ being a nonzero integer. It means that the number of particles transferred along the $x$ direction 
is equal to $\nu$, and particle distribution in real space is changed. 
Now if we assume the spectrum is gapped over the system on the cylinder, as shown in Fig.~\ref{fig:Laughlin}(c-1), the change of the boundary condition 
cannot change the particle distribution, because the shift of the allowed values of $k_y$ gives the particle distribution back to the 
original one (see Fig.~\ref{fig:Laughlin}(c-1)).
Therefore, the band structure for the cylinder system should have modes inside the gap. Suppose there is 
a branch with a positive dispersion in the gap (see Fig.~\ref{fig:Laughlin}(c-2)); then the number of particles changes by $-1$ by the 
change of $\phi$.
Instead, if there is 
a branch with a negative dispersion in the gap (see Fig.~\ref{fig:Laughlin}(c-3)), the number of particles changes by $+1$.
Note that such branches in the gap should be localized at edges of the cylinder, because the bulk band is assumed to have a gap.
Thus, to realize the nonzero change of $P_x$ (i.e. $\Delta P_x=\nu\neq 0$), the left end of the cylinder should have $\nu$ edge modes
with positive velocity ($\frac{\partial \omega}{\partial k_y}>0$) and the right end should have $\nu$ branches of edge modes with negative velocity ($\frac{\partial \omega}{\partial k_y}<0$) , 
as illustrated in Fig.~\ref{fig:Laughlin}(d). 
This physically means that for a two-dimensional system in an open geometry, there should be chiral edge modes, going along the system 
edge in  a clockwise way, and the number of branches of edge modes is $\nu$ (see Fig.~\ref{fig:Laughlin}(e)). When $\nu$ is negative, it means that the number of counterclockwise branches of edge modes
is $|\nu|=-\nu$. 

We emphasize that the discussions in Appendices~\ref{gauge invariance of Berry curvature}, \ref{quantization of the Chern number} and \ref{bulk-edge correspondence} can be applied both to discrete systems and to continuum systems including our system in the main text. 
In the discussions in Appendices~\ref{gauge invariance of Berry curvature}, \ref{quantization of the Chern number} and \ref{bulk-edge correspondence}, we did not assume the system to be discrete.

\begin{figure}
\includegraphics[width=9cm]{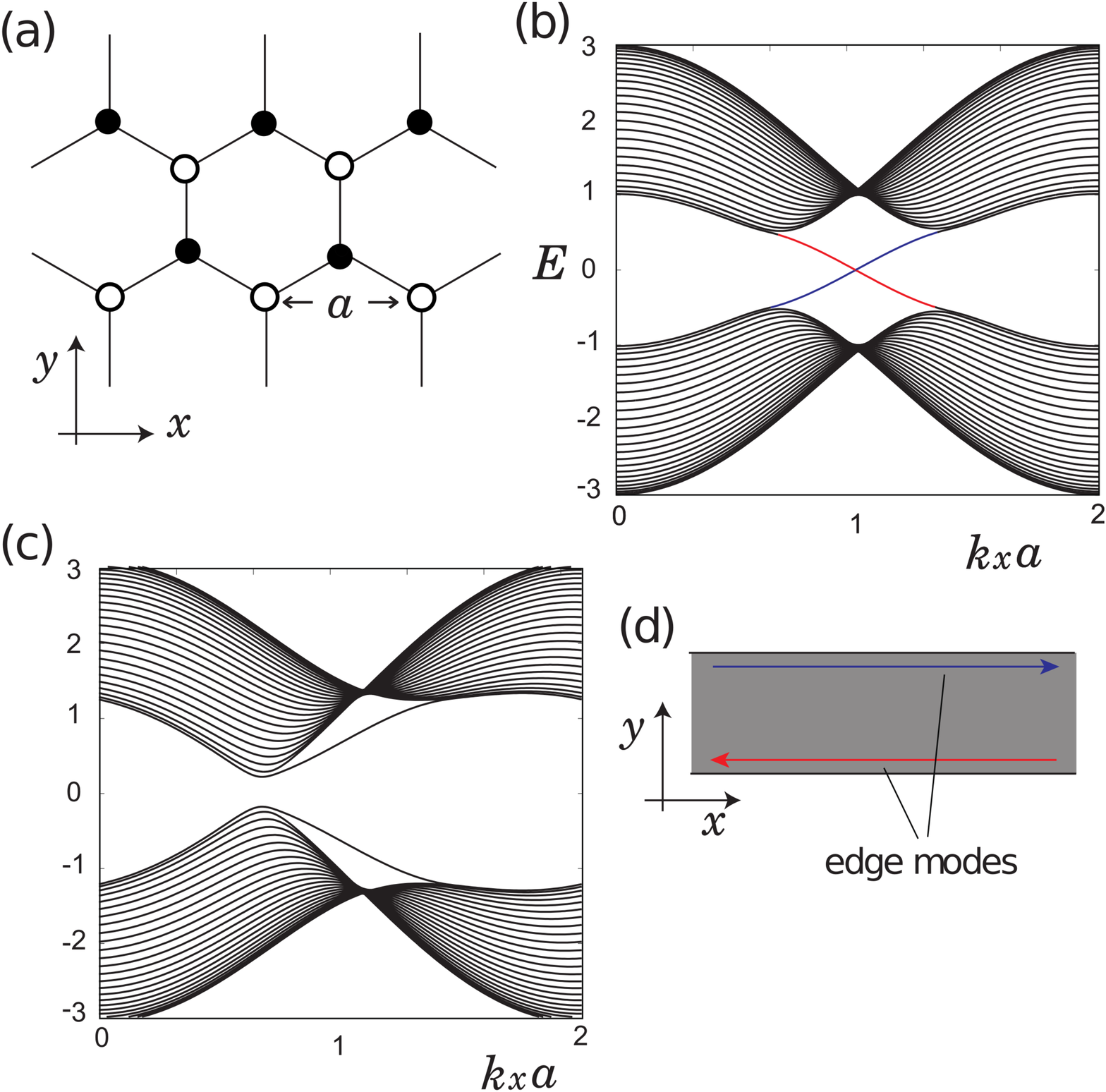}
\caption{\label{fig:Haldane} Haldane model and its band structures. (a) Honeycomb lattice. The unit cell consists of 
two sublattice sites, A and B, represented by solid and open circles, respectively. (b)(c) 
Band structures with zigzag-edge ribbon geometry
for (b) $t_1=1$, $t_2$=0.1, $M=0.0166$, $\varphi=1.57$
and (c) $t_1=1$, $t_2=0.1$, $M=0.7$, $\varphi=1.57$. The Chern number for the lower band is (b) $\nu=1$ and (c) $\nu=0$. (d) Schematic picture of 
edge modes in (b). The edge modes shown in blue and red lines correspond to those in (b) with the 
same colors.}
\end{figure}

One can numerically demonstrate this bulk-edge correspondence in various models. As an example, 
we show it for the Haldane model \cite{Haldane} on the honeycomb lattice, one of the well-known models for 
the quantum Hall systems. Here, we show existence of the topological chiral edge modes when
the Chern number is nonzero.
It is a tight-binding model on the honeycomb lattice (see Fig.~\ref{fig:Haldane}),
with the Hamiltonian  
\begin{equation}
H_{{\rm Haldane}}
=t_1\sum_{\langle i,j\rangle}c_{i}^{\dagger}c_j
+t_2\sum_{\langle\langle i,j\rangle \rangle}e^{-i\mu_{ij}\phi}c_{i}^{\dagger}
c_{j}+M\sum_{i}\xi_ic_i^{\dagger}c_i,
\label{eq:Haldane}
\end{equation}
where $c_i$ and $c_i^{\dagger}$ are annihilation and creation operators of particles at the site $i$, 
respectively. 
In the summations, $\langle i,j\rangle$ represents any nearest-neighbor pairs of sites $i$ and $j$, and
$\langle\langle i,j\rangle \rangle$ does any next nearest-neighbor pairs of sites $i$ and $j$.
Particles can be fermions or bosons, and the following results are the same for both cases.
For simplicity in explanation, fermions are assumed here. 
In Eq.~(\ref{eq:Haldane}), $t_1$, $t_2$, $M$ and $\phi$ are real, 
and $\mu_{ij}={\rm sgn}(\hat{\bm{d}}_{1}\times\hat{\bm{d}}_{2})_{z}=\pm 1$, 
where $\hat{\bm{d}}_{1}$ and $\hat{\bm{d}}_{2}$ are unit vectors along
the two nearest-neighbor bonds connecting between next-nearest neighbor pairs $i$ and $j$. 
Thus the 
next-nearest neighbor hopping $t_2 e^{\pm i\phi}$ is complex, with its
 phase being  $e^{i\phi}$
($e^{-i\phi}$) for a clockwise (counterclockwise) hopping in the hexagonal plaquette. $\xi_i$ represents a staggered on-site potential,
and takes values $\pm 1$ depending on the $i$-th sites being
in the A or B sublattices, respectively. 

In this model, phases with $\nu=1$, $\nu=0$ and $\nu=-1$ are realized by 
changing the parameters, when the Fermi energy is set to be $E_F=0$. In Figs.~\ref{fig:Haldane}(b) 
and (c) we show our numerical results of band structure calculation for a ribbon geometry; namely, 
the system is infinitely long in the $x$ direction and has a finite width in the $y$ direction,
and the edges are zigzag edges. In (b) and (c) the parameter values 
are
(b) $t_1=1$, $t_2$=0.1, $M=0.0166$, $\varphi=1.57$
and (c) $t_1=1$, $t_2=0.1$, $M=0.7$, $\varphi=1.57$, which yields the Chern number
to be (b) $\nu=1$ and (c) $\nu=0$. While the bulk bands have a gap around $E_F=0$ in 
both cases, there are two branches inside the gap in (b), while there is no branch in the gap in (c).
By analyzing the distribution of these in-gap branches, one can verify that one branch is 
localized at the lower edge while the other is at the upper edge, and they constitute 
clockwise edge modes; they are nothing but the topological edge modes expected from the
Chern number $\nu=1$, as schematically shown in Fig.~\ref{fig:Haldane}(d).

\section{\label{centrifugal}Effect of the centrifugal force}
In the main text, we neglected the effect of the centrifugal force. 
In this appendix we evaluate the condition to safely neglect it. 
The equation of motion with the centrifugal force term is written as
\begin{eqnarray}
&&\ddot{\mbox{\boldmath $u$}}=c_t^2 \mbox{\boldmath $\nabla$}^2\mbox{\boldmath $u$}+(c_l^2 - c_t^2)\mbox{\boldmath $\nabla$}(\mbox{\boldmath $\nabla$}\cdot\mbox{\boldmath $u$})+2\dot{\mbox{\boldmath $u$}} \times \mbox{\boldmath $\Omega$}\nonumber\\
&&\ \ 
-\mbox{\boldmath $\Omega$}\times(\mbox{\boldmath $\Omega$}\times\mbox{\boldmath $r$}).
\label{EOM+c}
\end{eqnarray}
Here we assume the system is a disk in the $xy$ plane with radius $R$, and the thickness along $z$ 
is sufficiently larger than the penetration depth of the SAW. Let the system rotate around the $z$ axis, and we take 
$\mbox{\boldmath $\Omega$}=(0,0,\Omega)$, $\bm{r}=(x,y,z)$, 
$\mbox{\boldmath $\rho$}=(x,y,0)$,  and $\rho=|\mbox{\boldmath $\rho$}|=\sqrt{x^2+y^2}$. 
Then, due to the centrifugal force, the system is stretched along the radial direction, parallel to 
$\bm{\rho}$. Thus, even in the absence of acoustic waves, the displacement $\bm{u}$ becomes
nonzero and is written as
\begin{equation}
\bm{u}=u(\rho)\hat{\bm{\rho}}
\end{equation}
where $\hat{\bm{\rho}}=\bm{\rho}/\rho$ is a unit vector along $\bm{\rho}$, $u(\rho)$ is a function of $\rho$ whose form is to be determined.
This displacement $\bm{u}(\bm{r})$ is a static solution of Eq.~(\ref{EOM+c}), leading to
\begin{eqnarray}
0=c_l^2 \frac{d^2u}{d\rho^2}+\Omega^2\rho
\label{EOM+c2}
\end{eqnarray}
with boundary conditions $u(\rho=0)=0$ and $\frac{du}{d\rho}(\rho=R)=0$, meaning that the system has no stress at the boundary. 
Its solution is 
\begin{eqnarray}
u=\frac{\Omega^2}{6c_l^2}(3R^2\rho-\rho^3)\ \ \ (0\leq\rho\leq R).
\label{u+c}
\end{eqnarray}
It is the static displacement in the radial direction, and it is positive as expected.
This leads to the expansion of the volume element with a ratio 
\begin{eqnarray}
\chi(\rho)\equiv 
\frac{u}{\rho}+\frac{du}{d\rho}=\frac{\Omega^2}{3c_l^2}(3R^2-2\rho^2)\ \ \ (0\leq\rho\leq R),
\label{du+c}
\end{eqnarray}
because the volume element increases from $dV\equiv \rho d\rho d\theta dz$ to
$dV'\equiv (\rho+u) d(\rho+u) d\theta dz\sim (1+\frac{u}{\rho}+\frac{du}{d\rho})dV$.
It has a maximum value $\chi(0)=\frac{\Omega^2R^2}{c_l^2}$ at $\rho=0$ and 
a minimum value $\chi(R)=\frac{\Omega^2R^2}{3c_l^2}$ at $\rho=R$. 
This expansion gives rise to a decrease of the mass density of the medium by a ratio $(1+\chi)^{-1}$, 
and the velocities of the
acoustic waves will increase by a ratio
\begin{equation}
(1+\chi)^{1/2}\sim 1+\frac{1}{2}\chi.
\end{equation}
Therefore, the frequency of the Dirac point becomes $(1+\chi(\rho)/2)\omega_K^{(D)}$, 
and it depends on $\rho$.
When this spatial variation of the frequency is much smaller than 
the size of the gap by the Coriolis force, $\omega_G$, one can safely ignore the centrifugal force. Therefore
its condition is given by 
\begin{equation}
\frac{1}{2}(\chi(0)-\chi(R))\omega_K^{(D)}= \frac{\Omega^2R^2}{3c_l^2}\omega_K^{(D)} \ll \omega_G.
\end{equation}
If we substitute the formula for $\omega_G$ in Eq.~(\ref{omegaG}), we get
\begin{eqnarray}
&&\frac{\Omega^2R^2}{3c_l^2}\omega_K^{(D)}  \ll \frac{9\sqrt{3}a^2 d}{N} \omega_K^{(D)} \Omega\\
&&\Rightarrow
\Omega \ll \frac{27\sqrt{3}a^2 dc_l^2}{N R^2}.
\end{eqnarray}

\end{document}